%% file: output.tex
\documentclass[journal, twocolumn]{IEEEtran}
\IEEEoverridecommandlockouts

\IEEEoverridecommandlockouts
\usepackage[dvips]{graphicx}
\usepackage[cmex10]{amsmath}
\usepackage{amsfonts,amssymb,bm}
\usepackage[tight,footnotesize]{subfigure}
\usepackage{fixltx2e}
\usepackage{dblfloatfix}
\usepackage{rotating}
\usepackage{multirow}
\usepackage{url}
\usepackage{cite}
\usepackage[linesnumbered,lined, ruled]{algorithm2e}

\usepackage{cite}
\usepackage{setspace}
\usepackage{footnote}
\usepackage[T1]{fontenc}
\usepackage{ae,aecompl}
\usepackage{epsfig}
\usepackage{multicol}
\usepackage{multirow}
\usepackage{subfigure}
\usepackage{times}
\usepackage{color}
\usepackage{comment}
\usepackage{amsthm}
\usepackage{caption}

\usepackage{colortbl}
\usepackage[official]{eurosym}

\SetCommentSty{mycommfont}
\newlength{\commentWidth}
\SetKwRepeat{Repeat}{repeat}{until}
\SetKwRepeat{Forever}{repeat}{forever}
\SetKwRepeat{On}{on}{end}
\SetNlSty{bfseries}{\color{black}}{}

\newtheorem{remark}{Remark}
\newtheorem{definition}{Definition}

\newcommand{\eg}[0]{\textit{e.g.},~}

\begin{document}

\vspace{-10pt}
\title{Fair Energy Allocation in Risk-aware Energy Communities}

%
\author{
\IEEEauthorblockN{Eleni Stai,~\IEEEmembership{Member,~IEEE},  Lesia Mitridati, \IEEEmembership{Member,~IEEE}, Ioannis Stavrakakis,~\IEEEmembership{Fellow,~IEEE}, Evangelia Kokolaki, Petros Tatoulis, Gabriela Hug,~\IEEEmembership{Senior Member,~IEEE}}

\thanks{E. Stai, P. Tatoulis and G. Hug are with the EEH - Power Systems Laboratory, ETH Z\"urich, Physikstrasse 3, 8092 Z\"urich, Switzerland.  L. Mitridati is with the Department of Wind \& Energy Systems, DTU. I. Stavrakakis is with the Department of Informatics and Telecommunications, National \& Kapodistrian University of Athens, 15784 Athens, Greece. E. Kokolaki is with the Hellenic Ministry of Environment and Energy, Mesogeion 119, 11526 Athens, Greece. E. Kokolaki’s work was carried out while she was with the National \& Kapodistrian University of Athens. Emails: elstai@ethz.ch, lemitri@dtu.dk, ioannis@di.uoa.gr, e.kokolaki@prv.ypeka.gr, petrost@student.ethz.ch, ghug@ethz.ch.}
}

\maketitle

\begin{abstract} 
This work introduces a decentralized mechanism for the fair and efficient allocation of limited renewable energy sources among consumers in an energy community. In the proposed non-cooperative game, the self-interested community members independently decide whether to compete or not for access to RESs during peak hours and shift their loads analogously. In the peak hours, a proportional allocation (PA) policy is used to allocate the limited RESs among the competitors. The existence of a Nash equilibrium (NE) or dominant strategies in this non-cooperative game is shown, and closed-form expressions of the renewable energy demand and social cost are derived. Moreover, a decentralized algorithm for choosing consumers' strategies that lie on NE states is designed. The work shows that the risk attitude of the consumers can have a significant impact on the deviation of the induced social cost from the optimal. Besides, the proposed decentralized mechanism with the PA policy is shown to attain a much lower social cost than one using the naive equal sharing policy.
\end{abstract}

\input{introduction}
\input{community}

\input{decentralized}

\input{centralized}

\input{algorithm}
\input{evaluation}

\input{conclusion}


%
\bibliographystyle{IEEEtran}
\vspace{-10pt}
\bibliography{output.bib}
\appendices
\input{appendix_new}
\clearpage


\end{document}

%% file: introduction.tex
\vspace{-0.18in}
\section{Introduction, Background \& Contributions}\label{sec:intro}

The large scale penetration of distributed, stochastic and non-dispatchable Renewable Energy Sources (RESs) has
triggered the need for energy management solutions in distribution systems 
\cite{muruganantham2017challenges,alam2019energy}.
In the meantime, growing environmental and societal awareness coupled with advances in communication and control technologies have allowed for a more active involvement of end-users in managing their energy consumption \cite{SCHWEIGER2020110359}. In the EU, recent regulatory changes, such as the 2019 Clean Energy for all Europeans package, and funding initiatives have placed a strong emphasis on renewable energy communities to enable local consumption and citizen-owned renewable energy projects \cite{EUpackage}.
In this context, energy communities, which coordinate the operation of distributed energy resources production and consumption, have become a viable and efficient solution to facilitate the integration of RESs into distribution grids, provide services to the grid and reduce procurement costs for consumers \cite{abada2020viability,charbonnier2022coordination}.



Several works in the literature 
have shown the benefits of energy communities to reduce consumers' costs and increase energy justice by focusing on peer-to-peer (P2P) energy trading mechanisms \cite{sousa2019peer,khan2023electricity}. However, the development of these energy communities with consumer-owned RESs may be limited due to high investment costs \cite{rodrigues2020battery}. In contrast, recent regulatory changes have provided an unprecedented opportunity for the development of P2P energy sharing mechanisms, in which the community-owned RESs are allocated efficiently and fairly among the consumers \cite{roberts2020power}. Various works in the literature showed the potential economic benefits of these energy sharing communities, both for individual consumers and the community as a whole \cite{lowitzsch2019investing,minuto2022energy, jia16}. For instance, the authors in \cite{jia16} show the effectiveness of community-owned RESs and storage versus a decentralized ownership using non-cooperative game theoretical tools. 

The design of fair and efficient mechanisms to share limited community-owned resources among consumers who have equal claim to them but different levels of demand is essential to ensure the large-scale development of these energy sharing communities \cite{moret2018energy,khan2023electricity}.
As highlighted in \cite{young1995equity}, due to the subjective nature of \textit{fairness}, various well-established notions of fairness, such as proportional and egalitarian, have been introduced in the literature and no allocation policy is universally accepted as "the most fair".
Additionally, different allocation policies satisfying one notion of fairness or another, may result in different levels of efficiency and stability. The work in \cite{bertsimas2011price} showed that allocation policies satisfying the notion of proportional fairness, such as the well-known \textit{proportional allocation (PA)} policy, may provide substantially higher efficiency and a lower "cost of fairness" than other axiomatically justified notions of fairness (e.g., egalitarian) by being more considerate to ``strong players'', i.e., consumers with high demand. Furthermore, the PA provides a trade-off between efficiency and fairness, since proportional fairness has been shown to be both Pareto optimal and a Nash bargaining solution \cite{boche2009nash}.
However, the PA may lack stability, as ``weak players'', i.e., consumers with low demand, may continuously change strategies to improve their allocation \cite{kulmala2021comparing}.
On the other hand, the well-established \textit{equal sharing (ES)} allocation policy, which satisfies an egalitarian notion of fairness, is known to provide greater stability than the PA since it allows small players to be fully satisfied and prevents strong players from obtaining more resources than other players. Yet, ES may result in highly inefficient and wasteful utilization of energy resources. In the context of demand response programs (DRPs) and the allocation of renewable resources, this is a major limitation to the application of ES.

Additionally, in order to be beneficial for the overall system, the interactions between prosumers in energy sharing communities and the grid must be coordinated, either through dynamic price signals, or organized local energy and flexibility markets \cite{pinson2014benefits,wheeler2023local,morstyn2021multiscale}.
Dynamic price signals, such as time of use (TOU) tariffs, which reflect wholesale energy prices and grid tariffs, can be implemented in a fully decentralized manner. This has made them more desirable in practice as a decentralized implementation is scalable and protects the privacy of consumers\cite{maharjan16,jacquot18}.
In the literature, a focus has been placed on designing efficient price signals to incentivize self-interested consumers to independently schedule their flexible loads in order to reduce their energy procurement costs and provide services to the grid (e.g. load shifting and peak load reduction) \cite{ibars2010distributed,Caron10,Joe_Wong12,Lulu2022}. Additional works studied the loss of efficiency arising from the self-interested behavior of consumers in these decentralized approaches compared to centralized scheduling approaches using non-cooperative game theoretical tools. The authors in \cite{ma2011decentralized} have shown the efficiency of price signals for an infinite population of consumers with identical technical characteristics and preferences. Similarly, the authors in \cite{chakraborty2017distributed} study the design of optimal price signals for deferral loads and derive bounds on the loss of efficiency. In addition, in \cite{shilov2021generalized}, the authors numerically illustrate the potential loss of efficiency of decentralized mechanisms in energy communities with consumers who have heterogeneous preferences as well as its impact on the grid under different pricing schemes. However, the aforementioned works fail to analyze the loss of efficiency resulting from the self-interested behavior of consumers with heterogeneous preferences that compete for multiple energy resources. Furthermore, a gap remains in analyzing the interactions between these price signals and energy sharing mechanisms in energy communities. 

This paper differentiates itself from the existing literature by addressing the aforementioned research gaps.
The framework for the energy sharing community considered is related to the multi-energy energy communities in \cite{maharjanuser,mitridati2021design}, and extends the preliminary work in \cite{stai2022}.
More precisely, we consider the interactions between self-interested consumers with heterogeneous flexibility preferences in an energy sharing community, subject to (i) a decentralized DRP, in which an energy retailer defines TOU tariffs to incentivize them to schedule their flexible loads across two time intervals and reduce grid imports during peak hours; and (ii) an energy sharing mechanism (ESM) to incentivize consumers to \textit{efficiently} and \textit{fairly} utilize the available resources.
The problem studied in this paper can model a wealth of resource allocation problems with a ternary cost structure beyond smart grids, such as the case of bandwidth or parking resources \cite{kokolaki2013}.

Given the aforementioned state-of-the-art, the contributions of this paper are the following:\\
$\bullet$ Firstly, we introduce a novel Decentralized Energy Sharing Mechanism (D-ESM) for an energy community, in which each consumer independently schedules its daily flexible loads at different time intervals, based on the availability and price of the different energy resources. In the proposed D-ESM, a wide range of consumers with heterogeneous preferences (namely daily energy demand and risk attitude) compete for access to multiple energy resources across two time intervals. This provides a novel application for the PA policy for multiple energy sources and time intervals.\\
$\bullet$ Secondly, we formulate the Centralized Energy Sharing Mechanism (C-ESM), in which a community manager centrally schedules flexible loads and allocates available resources based on the PA policy, as a linear optimization problem, and derive analytical solutions to it. This provides a benchmark against which to compare the efficiency of the D-ESM. \\
$\bullet$ Thirdly, we model and analyse the interactions among self-interested consumers participating in the proposed D-ESM, using non-cooperative game-theoretical tools. We introduce a novel game formulation of the proposed D-ESM, provide closed-form expressions of the stable operational points, i.e., the Nash Equilibrium (NE), and quantify the loss of efficiency due to the self-interested behavior of consumers compared to the C-ESM, via the Price of Anarchy (PoA) metric.\\
$\bullet$ Fourthly, we provide a novel iterative algorithm which prescribes the consumers' load schedules in a fully distributed and uncoordinated manner, so that they coincide with those prescribed by a NE. The proposed algorithm allows consumers to participate in the D-ESM without revealing privacy-sensitive information such as their individual loads and constraints. \\
$\bullet$ Finally, we provide thorough numerical analysis and comparisons between the proposed C-ESM and D-ESM, with emphasis on the Price-of-Anarchy (PoA) metric. We further compare the efficiency of the proposed D-ESM with the PA policy to that of a D-ESM based on an ES allocation policy. Additional fairness properties introduced via the distributed algorithm's design are studied.

The rest of the paper is organized as follows. 
In Section \ref{sec:esc}, we introduce the energy sharing community, the proposed D-ESM as well as the game-theoretic modeling approach of the interactions among consumers in the D-ESM. In Section \ref{sec:gameanalysis}, we study the NE mixed strategies under different parameter values. In Section \ref{sec:coordinated}, we investigate the solution via a C-ESM and in Section \ref{sec:efficiency}, we study the efficiency of the proposed D-ESM compared to the C-ESM. Section \ref{sec:algorithms} provides a distributed, uncoordinated algorithm with which the players can choose NE mixed strategies. Section \ref{sec:eval}, presents the numerical evaluations and comparisons. Finally, Section \ref{sec:conclusions} concludes the paper. 

%% file: community.tex
\vspace{-0.1in}
\section{Decentralized Energy Sharing Mechanism} \label{sec:esc}

\subsection{Energy Sharing Community} \label{sec:community}

The energy community consists of $N$ consumers, indexed by $i \in \mathcal{N}= \{1,...,N\}$, who have access to multiple energy sources in order to cover their flexible loads.

\subsubsection{Energy Sources}\label{sec:energy_sources}

We consider that the energy community has access to two distinct types of energy sources, namely \textit{local production} from community-owned RESs, and \textit{imports} from the distribution grid. We consider that the local RESs production is available only during daytime (e.g., PV panels), with a limited capacity $\mathcal{RE}>0$, whereas the community's imports from the grid are unlimited. Therefore, during nighttime the community's aggregate load is fully covered by imports from the grid, and, during daytime if the community's aggregate load exceeds the available RES capacity, the remainder is covered by imports from the grid.

Production from the community-owned RESs is priced by the community manager at a constant low tariff $c^{RES}$ (in units per energy), whereas imports from the grid are priced by an energy retailer using TOU tariffs, typically for daytime and nighttime consumption. We define the daytime and nighttime tariffs with respect to the RESs tariff, as $c^{grid,d}= \gamma c^{RES}$ and $c^{grid,n} = \beta c^{RES}$, respectively, with $\gamma>\beta>1$.
These TOU tariffs reflect the sum of energy prices and grid tariffs and are designed to incentivize consumers to shift their flexible loads from daytime to nighttime to reduce energy production costs and congestion during peak hours. In addition, the low cost of the local RESs production promotes self-consumption within the community and reduction of grid imports. We assume that the energy source-related parameters $\Omega = \{\mathcal{RE}, c^{RES}, \beta, \gamma\}$ are perfectly known by all consumers in the community at the beginning of the day.

\subsubsection{Consumers Preferences}

The consumers have a broad range of flexible loads, namely, (i) shiftable appliances (e.g. washing machines) that do not need to be scheduled every day, (ii) batteries or electric vehicles (EVs) with flexible state-of-charge requirements at the end of the day, and (iii) thermostatically controlled loads (e.g., water heater, heat pumps) with flexible set-points. The level of consumption and the time-schedule of these loads are flexible. For instance, an EV owner has a daily inflexible load required to cover her daytime transportation needs, and a daily flexible load, representing the additional energy to achieve a desired state-of-charge by the end of the day.
However, once scheduled, these loads cannot be interrupted or shifted to another time interval.
As a result, consumers whose daily flexible loads are scheduled during daytime incur the risk of paying for high-priced imports from the grid if the community's aggregate daytime energy demand exceeds the available local RES production. When scheduling their daily flexible loads across different time intervals, consumers wish to achieve a trade-off between their desired daily energy consumption, and the financial risks incurred. And, risk-averse consumers may choose to reduce their daily energy consumption if they are scheduled during daytime, to mitigate the financial risks incurred. For instance, if scheduled during nighttime, a risk-averse EV owner may prefer to consume enough energy to fully charge her EV by the end of the day, whereas, if scheduled during daytime, she may prefer to consume a smaller amount of energy in order to charge her EV at e.g., $75\%$ by the end of the day.

The risk attitude and daily energy consumption preferences of each consumer $i \in \mathcal{N}$ in the community can be represented by her type $\vartheta_i \in \Theta = \{1,...M\}$. The type accounts for consumer's (i) daily flexible load $U_{\vartheta_i}>0$ (in energy unit); and (ii) risk-aversion degree $\mu_{\vartheta_i} \in [0,1]$, representing the share of her daily flexible load that she is willing to consume if scheduled during daytime. 

With this parametric representation of the consumers' flexibility preferences, if the daily flexible load of a consumer $i$ of type $\vartheta_i$ is scheduled during daytime, her daytime energy demand is $E_{\vartheta_i} = \mu_{\vartheta_i} U_{\vartheta_i}$ (and the remainder of her daily flexible load $(1-\mu_{\vartheta_i})U_{\vartheta_i}$ is deferred to the following day), whereas, if her daily flexible load is scheduled during nighttime, her nighttime demand is $U_{\vartheta_i}$. Therefore, $\mu_{\vartheta_i}=1$ represents a risk-seeking consumer, and $\mu_{\vartheta_i} < 1$ a risk-conservative consumer. 

At the beginning of each day, each consumer knows her own flexibility preferences and type, but this information is considered private. We assume that the community manager and consumers in the community only know the probability distribution $\bm{r}=[r_1,...,r_{M}]^T$ over the consumers types $\Theta$, where $0 \leq r_{\vartheta} \leq 1$ is the probability that a consumer in the community is of type $\vartheta \in \Theta$. Furthermore, the consumers' preferences, and therefore their types, can vary from day to day. Since this paper studies a single scheduling day, the daily time indexes are omitted. 

Following the law of large numbers, the number of consumers of type $\vartheta \in \Theta$ can be approximated as $r_\vartheta \cdot N$. Thus, based on the above, the maximum daytime energy demand of the community, i.e., if the daily flexible loads of all consumers are scheduled during daytime is 
\vspace{-0.08in}
\begin{align}
 D^{Total} = N  \sum_{\vartheta \in \Theta} r_{\vartheta}~E_{\vartheta}.  
\end{align}
\vspace{-0.02in}
For notational simplicity, in the remainder of the paper, we introduce $\varepsilon_{\vartheta_i}=\frac{1}{\mu_{\vartheta_i}}$, such that $U_{\vartheta_i}=\varepsilon_{\vartheta_i} \cdot E_{\vartheta_i}$. Thus, $\varepsilon_{\vartheta_i}=1$ represents a risk-seeking consumer $i$, and $\varepsilon_{\vartheta_i}>1$ a risk-conservative consumer. Finally, we assume without loss of generality that $E_1\leq E_2 \leq ...\leq E_{M}$.

\vspace{-0.15in}
\subsection{Decentralized Energy Sharing Mechanism (D-ESM)}

The problem faced by the energy sharing community is to schedule the daily flexible loads of all consumers across the different TOU intervals and to allocate the different energy sources among them within each TOU interval. The role of the community manager is to design a mechanism that optimally coordinates the interactions among consumers in the community towards desirable outcomes, namely: (i) minimizing the social cost for the community as a whole, and (ii) sharing the community-owned assets among the consumers fairly. We introduce below the proposed decentralized energy sharing mechanism (D-ESM) for this energy sharing community.

\subsubsection{Load Scheduling}

In the proposed D-ESM, each consumer independently schedules her own daily flexible loads across the different TOU intervals, at the beginning of the day, in order to maximize her own utility under the set energy source allocation and payment policies. In contrast, in a Centralized ESM (C-ESM), the community manager would schedule the daily flexible loads of all consumers across the different TOU intervals in order to minimize the social cost of the community as a whole under the set energy source allocation and payment policies (see Section \ref{sec:coordinated}). As implementing this centralized approach would require for the community manager to have information on each consumer's preferences, it can only be considered as an ideal benchmark against which to compare the efficiency of the proposed D-ESM.

In this paper, we study \textit{mixed strategies} of consumer types. A \textit{mixed strategy} is a probability distribution $\mathbf{p}_{\vartheta}=[p_{\vartheta}^d, p_{\vartheta}^n]^T$, with $p_{\vartheta}^d \in [0,1]$ denoting the probability that a consumer of type $\vartheta \in \Theta$ schedules her daily flexible load during daytime, and $p_{\vartheta}^n \in [0,1]$ during nighttime. At the beginning of the day a consumer $i$ determines her mixed strategy based on her type $\vartheta_i \in \Theta$, $\mathbf{p}_{\vartheta_i}$. Then she schedules her daily flexible loads either in daytime or in nighttime with probabilities $p_{\vartheta}^d$, $p_{\vartheta}^n$, correspondingly. Let also $\bm{p}$ be the collection of mixed strategies of all consumers, i.e., $\bm{p}= \{\bm{p}_{\vartheta_i}\}_{i \in \mathcal{N}}$. 

 
\subsubsection{Energy Source Allocation and Payment Policies} \label{sec:policies}
 
Once the daily flexible loads of all consumers have been scheduled, the community manager must allocate the available energy sources at each TOU interval (daytime or nighttime) among them. During nighttime, all scheduled loads are covered by grid imports since this is the sole available energy source for this TOU interval. During daytime, the community manager allocates in priority the local RESs production to cover the scheduled daytime loads, in order to maximize local consumption from the community and reduce energy costs. However, if the expected aggregate daytime energy demand exceeds the available local RESs production, the community manager must share this limited resource among those consumers with loads scheduled during daytime. This raises the challenging issue of allocating fairly a limited resource among users with equal claims to it.

In order to ensure a notion of fairness among community members, the community manager allocates to each consumer $i$ of type $\vartheta_i$ a share of the local RESs production proportional to her daytime load schedule. As a result, under this PA policy, the local RESs production allocated to a consumer whose daily flexible load is scheduled during daytime is

\vspace{-0.1in}
\begin{small}
\begin{eqnarray}
res^{PA}_{\vartheta_i}(\mathbf{p}) &=& \frac{E_{\vartheta_i}}{\max(\mathcal{RE} , D^d(\mathbf{p}))}\mathcal{RE},
\label{eq:prop_alloc_energy}
\end{eqnarray}
\end{small}
\vspace{-0.1in}

\noindent where $D^d(\mathbf{p})$ denotes the expected aggregate daytime demand of the community.

Each consumer $i$ of type $\vartheta_i$ must then pay for the different energy sources covering her scheduled load at each TOU interval, ensuring budget balance of the proposed mechanism.

\vspace{-0.1in}
\subsection{Non-cooperative Game Formulation} \label{sec:game_def}

Based on the proposed D-ESM framework, if a consumer schedules her daily flexible load during daytime, she competes with other consumers to use the limited local RESs production and incurs a financial risk. This competition among the consumers participating in the proposed D-ESM (for one single day) can be modeled as an Energy Sharing Game (ESG), as defined bellow.


\vspace{-0.05in}
\begin{definition}\label{def:energy_source_game}
An \emph{Energy Sharing Game (ESG)} is a single-shot noncooperative game, defined by the tuple
\\$\Gamma=(\mathcal{N}, \{\mathcal{P}_{\vartheta_i}\}_{i\in\mathcal{N}}, \{\upsilon_{\vartheta_i}\}_{i\in \mathcal{N}})$, where:
\begin{itemize}
    \item $\mathcal{N}=\{1,...,N\}$ is the set of players, i.e., the consumers in the energy sharing community.
    \item $\mathcal{P}_{\vartheta_i} = \{ \mathbf{p}_{\vartheta_i} | \mathbf{p}_{\vartheta_i} : A_i \in \mathcal{A} \rightarrow p^{A_i}_{\vartheta_i} \in \mathbb{R}^+ , \text{ with } \sum_{A_i \in \mathcal{A}} p^{A_i}_{\vartheta_i} = 1 \}$ is the set of mixed strategies of player $i$ of type $\vartheta_i$ over the set of pure strategies $\mathcal{A}=\{d,n\}$, consisting of the choices to schedule her daily flexible load during daytime ($A_i = d$) or during nighttime ($A_i = n$). Therefore, each consumer $i$ of type $\vartheta_i$ with a mixed strategy $\mathbf{p}_{\vartheta_i}$, plays this game by randomly selecting an action $A_i \in \mathcal{A}$ with probability $p^{A_i}_{\vartheta_i}$\footnote{Note that a \textit{pure strategy} is a special case of a mixed strategy where one action has a probability equal to 1 (and the remaining have 0).}.
    \item $\upsilon_{\vartheta_i} : A_i \in \mathcal{A} \rightarrow  \upsilon^{A_i}_{\vartheta_i}$ is the payoff function of a consumer $i$ of type $\vartheta_i$ over the set of pure strategies $\mathcal{A}$. The cost of a consumer $i$ of type $\vartheta_i$ who plays the pure strategy $A_i = d$, is 
    \begin{small}
    \begin{align}
    \upsilon^{d}_{\vartheta_i} = c^{RES} res_{\vartheta_i}^{PA}(\mathbf{p}) + c^{grid,d} (E_{\vartheta_i}-res_{\vartheta_i}^{PA}(\mathbf{p})),
    \label{eq:RES_cost}
    \end{align}
    \end{small}
    \hspace{-0.1in} and depends on the strategy profile $\bm{p}$ of all consumers via the community's expected aggregate daytime energy demand $D^d(\mathbf{p})$. The cost of a consumer who plays the pure strategy $A_i = n$ is
    \begin{small}
    \begin{align}\upsilon^{n}_{\vartheta_i} = U_{\vartheta_i}  c^{grid,n},
    \label{eq:nonRES_cost}
    \end{align}
    \end{small}
    \hspace{-0.1in} and is independent on other consumers' mixed strategies.
    Before making their decisions all players have perfect knowledge of the energy sources parameters in the set $\Omega$ and their own preferences and type, and have prior knowledge on the probability distribution $\bm{r}$ over the other consumers' types.
\end{itemize}
\end{definition}

A consumer of type $\vartheta \in \Theta$ repeatedly playing the mixed strategy $\bm{p}_{\vartheta}$ over multiple instances of the ESG 
would have an \textit{expected} daytime and nighttime energy demand equal to $D_{\vartheta}^{d} = p_{\vartheta}^{d} E_{\vartheta}$ and $D_{\vartheta}^{n} = p_{\vartheta}^{n} U_{\vartheta}$, respectively. Therefore, the mixed strategy of a consumer $i$ of type $\vartheta_i$ can alternatively be interpreted as splitting her daily flexible loads between daytime and nighttime, such that her daytime load schedule is equal to $D_{\vartheta_i}^{d}$, and their nighttime load schedule is equal to $D_{\vartheta_i}^{n}$. 
With these notations, the \textit{expected} aggregate daytime and nighttime energy demands of the community are respectively expressed as

\vspace{-0.1in}
\begin{small}
\begin{subequations}
\begin{align}
    & D^d (\mathbf{p}) = N\sum_{\vartheta\in \Theta} r_{\vartheta }~p^d_{\vartheta}~E_{\vartheta,}\label{eq:demand} \\
     & D^n (\mathbf{p}) = N\sum_{\vartheta\in \Theta} r_{\vartheta }~p^n_{\vartheta}~U_{\vartheta}\label{eq:demand_n}.
 \end{align}
\end{subequations}
\end{small}
\vspace{-0.15in}

%% file: decentralized.tex
\vspace{-0.1in}\section{Analysis of the Decentralized Energy Sharing Mechanism}\label{sec:gameanalysis}


In this section, we study analytically the uncoordinated decisions of the self-interested consumers participating in the proposed D-ESM. In the following, we study the conditions on the parameter values for the existence of dominant strategies and mixed-strategy NE under the proposed PA and payment policies, and provide closed-form formulations of these equilibrium states, i.e., ranges on the values of the vector of mixed strategies at NE, denoted as $\mathbf{p^{NE}}$ and an analytical expression on the value of the expected aggregate daytime energy demand. The proofs of the theoretical results presented below are available in the {Appendix \ref{sec:proofsESSG}} of \cite{arxiv_version}.

First, we recall that, for a mixed-strategy NE to exist, the expected costs of each consumer for all pure strategies in the support of the mixed-strategy NE must be equal. Using the expressions of the costs in \eqref{eq:RES_cost} and \eqref{eq:nonRES_cost}, we obtain that the amount of RESs allocated to a consumer type $\vartheta \in \Theta$ at a NE must satisfy:

 \vspace{-0.1in} 
\small
\begin{equation}\label{eq:conditionEQ_extra_demand}
res^{NE}_{\vartheta}(\mathbf{p^{NE}}) = \frac{\gamma-\varepsilon_{\vartheta}\beta}{\gamma-1}E_{\vartheta},~ \forall \vartheta \in \Theta.
\end{equation}
\normalsize 
\vspace{-0.1in}

\noindent Thus, in the ESG, a mixed-strategy NE exists under the condition:

 \vspace{-0.1in} 
 \small
\begin{equation}\label{eq:condition_PA_NE}
res_{\vartheta}^{PA}(\mathbf{p}^{NE}) = res_{\vartheta}^{NE}(\mathbf{p}^{NE}), ~\forall \vartheta \in \Theta, \end{equation}
\normalsize 
\vspace{-0.15in}

\noindent where $res_{\vartheta}^{PA}(\mathbf{p}^{NE})$ is defined in \eqref{eq:prop_alloc_energy}. In the following analysis, we obtain the mixed-strategy NE competing probabilities $\mathbf{p}^{NE}$ by solving Equation \eqref{eq:condition_PA_NE}. We further distinguish cases with respect to the available RESs production, TOU tariffs, and consumers' types.

\subsubsection*{\textbf{Case $1$: $\bm{\mathcal{RE}}$ exceeds $\bm{D^{Total}}$}}

As consumers have knowledge of $\mathcal{RE}$ and $D^{Total}$, it is straightforward to show that the dominant-strategy for all consumers is to schedule their daily flexible loads during daytime. As a result, the competing probabilities that lead to equilibrium states are equal to $p_{\vartheta}^{d,NE} = 1$ for all consumer types $\vartheta \in \Theta$.

\subsubsection*{\textbf{Case $2$: $\bm{\mathcal{RE}}$ is lower than $\bm{D^{Total}}$}} 

In this case, the strategies of the consumers depend on their respective risk aversion degrees and the TOU tariffs. We define two complementary subsets of consumer types, depending on their risk aversion degrees: $\Sigma_1 = \Bigl\{\vartheta \in \Theta : \varepsilon_{\vartheta} \geq \gamma/\beta \Bigr\} \subset \Theta$, and $\Sigma_2 = \Bigl\{ \vartheta \in \Theta :1\leq \varepsilon_{\vartheta} < \gamma/\beta \Bigr\} \subset \Theta$.

Firstly, the dominant strategy for all consumers $i$ whose type $\vartheta_i$ is in the set $\Sigma_1$ is to schedule their daily flexible loads during daytime, i.e., to play the pure strategy $A_i = d$ with probability $p_{\vartheta_i}^{d,NE} = 1$. 

Secondly, the strategies of the consumers $i$ whose type $\vartheta_i$ is in the set $\Sigma_2$ depend on their daily flexible loads and risk-aversion degrees. We define two distinct subsets of consumer types in $\Sigma_2$: $\Sigma_{2,1} = \left\{ \vartheta \in \Sigma_2 : E_{\vartheta} > \mathcal{RE}\frac{(\gamma-1)}{(\gamma-\varepsilon_{\vartheta}\beta)} \right\}$ and $\Sigma_{2,2} = \left\{ \vartheta \in \Sigma_2 : E_{\vartheta} \leq \mathcal{RE}\frac{(\gamma-1)}{(\gamma-\varepsilon_{\vartheta}\beta)}\right\}$.


For consumers $i$ whose type $\vartheta_i$ is in the set $\Sigma_{2,1}$, the dominant strategy is to schedule their daily flexible loads during nighttime, i.e., to play the pure strategy $A_i=n$ with probability $p^{n,NE}_{\vartheta_i}=1$ and $A_i=d$ with probability $p^{d,NE}_{\vartheta_i}=0$.

For consumers $i$ whose type $\vartheta_i$ is in the set $\Sigma_{2,2}$, a mixed-strategy NE under the PA policy exists if and only if the following condition holds:

 \vspace{-0.1in} 
 \small
\begin{equation}\label{eq:relation_E_0_E_1_pa_ne_extra_demand}
\begin{split}
& \mathcal{RE}\frac{(\gamma-1)}{(\gamma-\varepsilon_{\vartheta}\beta)}-E_{\vartheta}  = \mathcal{RE}\frac{(\gamma-1)}{(\gamma-\varepsilon_{\tilde{\vartheta} }\beta)}-E_{\tilde{\vartheta}} , \ \forall \vartheta , \tilde{\vartheta} \in \Sigma_{2,2}.
\end{split}
 \end{equation}
\normalsize 
\vspace{-0.1in}

\noindent Assuming that all consumers of the same type play the same mixed strategy, the competing probabilities that lead to NE states lie in the range $p^{min}_{\vartheta} \leq p^{d,NE}_{\vartheta} \leq p^{max}_{\vartheta}$ for all consumer types $\vartheta \in \Sigma_{2,2}$ with:

 \vspace{-0.1in} 
 \footnotesize
   \begin{align}
& p^{max}_{\vartheta} = \min \left\{1,\frac{\frac{\mathcal{RE}(\gamma-1)}{(\gamma-\varepsilon_{\vartheta}\beta)}-D^{Total}_{\Sigma_1}}{N~ r_{\vartheta}~ E_{\vartheta}}\right\},
    \label{eq:plmaxbound} \\
&  p^{min}_{\vartheta}= \max \left\{0,\frac{\frac{\mathcal{RE}(\gamma-1)}{(\gamma-\varepsilon_{\vartheta}\beta)}-D^{Total}_{\Sigma_1 \bigcup \Sigma_{2,2}\setminus \{\vartheta\}}}{N~ r_{\vartheta}~E_{\vartheta}} \right\},\label{eq:plminbound}
  \end{align}
\normalsize  
where for any subset of consumer types $\mathcal{S} \subset \Theta$, $D^{Total}_{\mathcal{S}}$ represents the maximum aggregate daytime demand of consumers whose type is in $\mathcal{S}$, e.g., $D^{Total}_{\Sigma_1}=N\sum_{\theta \in \Sigma_1}r_{\theta} E_{\theta}$.

 As a result, the expected aggregate daytime demand, $D^{d,NE}$ at NE is

\vspace{-0.1in} 
\footnotesize
\begin{align}
& D^{d,NE} = D^{Total}_{\Sigma_1} \nonumber \\
& + \min \left\{ D^{Total}_{\Sigma_{2,2}} , \max \left\{\frac{N\left(\frac{\mathcal{RE}(\gamma-1)}{(\gamma-\varepsilon_{\vartheta}\beta)}-E_{\vartheta}-D^{Total}_{\Sigma_1}\right)}{(N-1)},0 \right\} \right\}. \label{eq:demand1_2c}
\end{align}
\normalsize 

\begin{remark} \label{rem:risk_degrees_relation}
Note that condition \eqref{eq:relation_E_0_E_1_pa_ne_extra_demand} can hold, and therefore a NE can exist, only if for any pair $\vartheta , \tilde{\vartheta} \in \Sigma_{2,2}$ such that $\vartheta \leq \tilde{\vartheta}$, it holds that $\varepsilon_{\vartheta} \leq \varepsilon_{\tilde{\vartheta}}$. Since by assumption, $E_{\vartheta} \leq E_{\tilde{\vartheta}}$, this means that consumers with lower daytime energy demand levels should be more risk-seeking than those with higher ones.
\end{remark}
\begin{remark}\label{rem:risk_seeking}In particular, if all consumers whose type is in $\Sigma_{2,2}$ are risk-seeking (i.e., $\varepsilon_{\vartheta}= 1, \forall \vartheta \in \Sigma_{2,2}$), a NE can only exist if  $E_{\vartheta}=E_{\tilde{\vartheta}} , \ \forall \vartheta , \tilde{\vartheta} \in \Sigma_{2,2}$.
\end{remark}



%% file: centralized.tex
\section{Centralized Energy Sharing Mechanism}\label{sec:coordinated}

In this section we study an ideal centralized scheduling problem, in which an energy community manager with perfect knowledge of the available energy sources and types of the consumers in the community, centrally schedules their daily flexible loads.

\subsection{Problem Formulation}


Based on the available information, the community manager aims at finding the optimal load schedule of each consumer type, which minimize the social cost of the community under the chosen PA and payment policy. The community's social cost $C^{PA}(\bm{p})$ can be expressed as a function of the \textit{expected} aggregate daytime energy demand ($D^d(\bm{p})$) and nighttime energy demand ($D^n(\bm{p})$) of the community (as defined in Section \ref{sec:game_def}), such that:

\vspace{-0.15in} 
\begin{small}
	\begin{align}
	C^{PA}(\bm{p}) &=  \min\{\mathcal{RE}, D^d(\bm{p})\} \cdot c^{RES} \nonumber \\
    &+ \max\{0,D^d(\bm{p})-\mathcal{RE}\} \cdot c^{grid,d} + D^n(\bm{p}) \cdot c^{grid,n},
\label{eq:social_cost_pa_extra_demand}
	\end{align}
\end{small}

\noindent where the probabilities $p^d_{\vartheta}$ and $p^n_{\vartheta}$ (as defined in Section \ref{sec:game_def}) can be interpreted as the proportion of consumers of type $\vartheta$ that the community manager schedules during daytime and nighttime, respectively.
Although this objective cost is non-convex, we observe that during daytime, for any expected aggregate load schedule, the community manager minimizes the cost from grid imports. Therefore, by introducing the optimization variable $D^{grid}$ representing the expected aggregate grid imports during daytime, we can write the community manager's optimal load scheduling problem under the PA policy as a linear optimization problem, as follows:

 \vspace{-0.1in} 
 \begin{small}
 \begin{subequations} \label{eq:social_cost_x_opt}
\begin{alignat}{2}
& \min_{\mathbf{p},D^{grid}} \ && c^{grid,d}  D^{grid} + c^{RES}  \left(N\sum_{\vartheta \in \Theta} r_{\vartheta} p_{\vartheta}^d E_{\vartheta} - D^{grid}\right) \nonumber \\
& \quad && + c^{grid,n}  N \sum_{\vartheta \in \Theta} r_{\vartheta} {p}_{\vartheta}^{n}U_{\vartheta} \label{eq:opt_1} \\
 & \text{s.t. } &&  p^{d}_{\vartheta} + p^{n}_{\vartheta} = 1 , \ \forall \vartheta \in \Theta, \label{eq:opt_2.1} \\
  & \quad && 0 \leq p^{d}_{\vartheta},  p^{n}_{\vartheta},  \ \forall \vartheta \in \Theta, \label{eq:opt_2.2} \\
 & \quad && D^{grid} \geq \mathcal{ER} - N \sum_{\vartheta \in \Theta} r_{\vartheta} {p}_{\vartheta}^{d} E_{\vartheta}, \label{eq:opt_3.1} \\
 & \quad && D^{grid} \geq 0. \label{eq:opt_3.2}
 \end{alignat}
 \end{subequations}
\end{small} \vspace{-0.1in}  

\noindent This optimization problem minimizes the social cost of the community \eqref{eq:opt_1}, subject to constraints on the daytime and nighttime probabilities \eqref{eq:opt_2.1}-\eqref{eq:opt_2.2}  as well as to lower bounds on the expected aggregate grid imports during daytime \eqref{eq:opt_3.1}-\eqref{eq:opt_3.2}.

\subsection{Solution Analysis} \label{sec:centralsol}

In the following, we provide insights and analytical formulations of the optimal solutions $\mathbf{p^{*}}$ of this centralized mechanism in different cases. The proofs are available in the {Appendix \ref{appendix:dual}} of \cite{arxiv_version}.
 
\subsubsection*{\textbf{Case $1$: $\bm{\mathcal{RE}}$ exceeds $\bm{D^{Total}}$}}

In this trivial case, the optimal solutions to the C-ESM is to schedule all consumers' daily flexible loads during daytime, such that $p^{d,*}_{\vartheta}=1$, $\forall \vartheta \in \Theta$, and the expected grid imports $D^{grid,*} =0$.

\subsubsection*{\textbf{Case $2$: $\bm{\mathcal{RE}}$ is lower than $\bm{D^{Total}}$}}

In this case, it is optimal for the centralized ESM to schedule loads during the day so that the total RES capacity is fully utilized. To perform the analysis, we use the two complementary subsets of consumer types, $\Sigma_1$ and $\Sigma_2$, as those are defined in Section \ref{sec:gameanalysis}. 

For all consumers whose type $\vartheta \in \Sigma_1$, it is optimal for the community to schedule them during daytime, such that $p^{d,*}_{\vartheta}=1 $. For the optimal load schedule of the remaining consumers whose type $\vartheta \in \Sigma_2$, we observe that the consumer types are scheduled during daytime in order of increasing risk aversion (i.e., decreasing $\varepsilon_\vartheta$), until the local RESs production is fully utilized. Therefore, the optimal competing probabilities for the consumers whose types are in $\Sigma_2=\{\tilde{\vartheta}^1, \tilde{\vartheta}^2, \dots, \tilde{\vartheta}^K \}$, can be expressed as:

 \vspace{-0.1in} 
 \footnotesize
\begin{align}
 & p^{d,*}_{\tilde{\vartheta}^k} = \max \Bigg\{ \min \Bigg\{ 1, \dfrac{\left( \mathcal{RE} - D^{Total}_{\Sigma_1} - N \sum_{i=1}^{k-1}r_{{\tilde{\vartheta}^i}} E_{{\tilde{\vartheta}^i}} p^{d,*}_{\tilde{\vartheta}^i}  \right)}{N r_{{\tilde{\vartheta}^k}} E_{{\tilde{\vartheta}^k} }}\Bigg\} , 0 \Bigg\} , \nonumber \\
&  \forall k \in \{1,...,K\},
\end{align}
\normalsize 
\vspace{-0.1in}  

\noindent where the consumer types in $\Sigma_2$ are ordered such that $\varepsilon_{\tilde{\vartheta}^1} \geq \varepsilon_{\tilde{\vartheta}^2} \geq ... \geq \varepsilon_{\tilde{\vartheta}^K}$.


\section{Efficiency Loss of D-ESM vs. C-ESM}\label{sec:efficiency}
The (in)efficiency of equilibrium strategies in the D-ESM compared to the optimal C-ESM solution is quantified by the Price of Anarchy (PoA) metric \cite{Koutsoupias09}, representing the ratio of the worst case social cost among all mixed strategy NE, denoted as $C^{PA,NE}_{WC}$, over the optimal minimum social cost of the C-ESM, such that:

 \vspace{-0.1in} 
 \small
\begin{align}
 \hspace{-5pt} \textit{PoA}  = \frac{C^{PA,NE}_{WC}}{C^{PA}(\mathbf{p^{^*}})}.
\label{eq:poa_pa}
\end{align}
\normalsize 
\vspace{-0.1in}  

First observe that $C^{PA}(\mathbf{p^{^*}})$ is uniquely determined for each particular case (Section \ref{sec:coordinated}). Now, in order to obtain $C^{PA,NE}_{WC}$ when there exist multiple possible NE, we can maximize the social cost $C^{PA}(\mathbf{p^{NE}})$ (Eq. \eqref{eq:social_cost_pa_extra_demand}) with respect to $\mathbf{p^{NE}}$.




%% file: algorithm.tex
\section{Distributed Algorithm to Obtain NE} \label{sec:algorithms}
\vspace{-0.05in}
In this section, we design a distributed, uncoordinated algorithm that computes consumers' mixed-strategies that lie on NE for the ESG when there does not exist a dominant-strategy for each consumer, i.e., for the consumers in the set $\Sigma_{2,2}$ of Case 2. Note that given their knowledge on the set $\Omega$ (Section \ref{sec:esc}), the consumers can know whether they have a dominant strategy and in such a case they can directly compute it. The proposed distributed iterative algorithm, Algorithm \ref{algorithm}, is based on a best response scheme and requires minimum information exchange among consumers. In particular, there is no need of central coordinator or direct communication channels between consumer pairs since the required information can be just broadcasted from the consumer that has performed the most recent computation to the remaining ones. 

The outer loop represents the algorithm's steps, and the inner loop iterates over all consumers who are randomly ordered in a list $\Sigma$ and at each iteration, they update their strategies. All consumers with a certain type share the same strategy in each algorithm's step. 
 Hence, if consumer $i$'s type has already been assigned a probability by another consumer of the same type in a previous iteration of the inner loop, consumer $i$ just retrieves this probability value (line \ref{line:retrievestrategy}), otherwise it computes the best response of its type to the types that have already played (lines \ref{line:PAconstr}-\ref{line:pacomputed}). Although, the inner loop practically computes consumer type strategies, it iterates over all consumers and not over all consumer types so as to allow for distributed operation; otherwise a central entity is needed to compute the consumer type strategies. 
As we aim to limit information exchange, the chosen strategies are not communicated. Instead the consumers update and broadcast three common variables, which encode this information:

\noindent $1)$ the variable $X_\Sigma$ that is equal to the total current daytime energy demand;\\
$2)$ the vector $EQT$ that indicates which consumer types have played in the previous iterations of an algorithm's step ($EQT(\vartheta)=1$ if type $\vartheta \in \Theta$ has played), and is re-initialized to $\bm{0_M}$ at the beginning of each outer loop;\\
$3)$ the vector $EQP$ that contains the current mixed strategies values for all consumer types and is updated each time a consumer type updates its strategy (line \ref{line:probincrease}).


Based on the values of these common variables, each consumer type $\vartheta \in \Theta$, in its turn, updates its strategy by minimizing its expected cost of energy (line \ref{line:bestresponse}), given by:

\vspace{-0.1in}
\begin{small}
\begin{align}
    &cost(p_{\vartheta}^{d})  \nonumber =\\& p_{\vartheta}^{d}\big[res_{\vartheta}^{PA}(\mathbf{p})\cdot c^{RES}+(E_{\vartheta}-res_{\vartheta}^{PA}(\mathbf{p})) \cdot\gamma \cdot c^{RES}\big]\nonumber \\ 
    & 
    + (1-p_{\vartheta}^{d}) \cdot U_{\vartheta}\cdot \beta \cdot c^{RES}. \label{eq:personalcost}
\end{align}
\end{small}

A limitation of the best response scheme is that the first consumer that plays at an algorithm's step can freely choose her daytime RES demand. In order to mitigate this effect, we introduce a \textit{capping system} at the inner loop, which multiplies the best response with a parameter $cap\in [0,1]$\footnote{$cap$ can be constant through the algorithm or drawn from a uniform distribution. This will be discussed in the numerical evaluations.} (line \ref{line:cap}), such that the adjusted response is

\vspace{-0.15in}
\begin{small}
\begin{align}\label{eq:cap}
    f^{cap}(p_{\vartheta}^{d}) &= cap \cdot p_{\vartheta}^{d}.
\end{align}
\end{small}
\vspace{-0.1in}

As a result, even after the completion of an algorithm's step, it is possible that the total available RES capacity has not been allocated. In this case, additional outer steps are needed in order to reach an equilibrium state. In practice, the algorithm continues until one of the two following conditions hold: (i) a NE is reached, which means that the players do not wish to change their actions unilaterally with respect to the previous step (line \ref{line:equilibrium}), or (ii) a maximum number of steps ($N_{step}$) is reached.

\begin{small}
\begin{algorithm}[t] 
    \SetAlgoLined
    \textbf{Input} 
    $N_{step}$: number of algorithm's steps\\
    \textbf{Output} $EQP$: Vector of NE mixed strategies for each consumer type; \\\
    \textbf{Initialization: \label{line:init}}\\
    $(p^d_{\vartheta}, p^n_{\vartheta}) \gets (0,1), ~\forall \vartheta  \in \Theta$ \;
    $X_\Sigma \gets N \sum_{ \vartheta \in \Sigma_1  } r_{\vartheta} E_{\vartheta}$, ~~$\Sigma \gets \Sigma_{2,2}$ \;
    $EQP \gets$ vector of size $M$ with zero entries for $\vartheta \in \Sigma$ and unary entries for $\vartheta \in \Sigma_{1}$\;
    \For{$step \gets 1$ \KwTo $N_{step}$ \label{line:external_round}}{
        $EQT \gets \mathbf{0}_M$\;
        $EQP_{old} \gets EQP$\;
        \For{each consumer $i \in \Sigma$}{
            \If{$EQT(\vartheta_i) = 1$}{
            Consumer $i$ retrieves  $p^d_{\vartheta_i}$ from $EQP(\vartheta_i)$\; \label{line:retrievestrategy}}
            \Else{
                {\small$res_{\vartheta_i}^{PA}(p_{\vartheta_i}^{d})\leftarrow \frac{E_{\vartheta_i}\cdot \mathcal{RE}}{X_\Sigma +(N-1) ~r_{\vartheta_i} p^{d}_{ \vartheta_i} E_{\vartheta_i} + E_{\vartheta_i}}$}\label{line:PAconstr} \;
               {\small$p_{\vartheta_i}^{d,*}\gets \,\underset{p^d_{\vartheta_i}} {\mathrm{arg \min}}\, cost(p_{\vartheta_i}^{d})$}\label{line:bestresponse}, from \eqref{eq:personalcost}\;
                {\small$p_{\vartheta_i}^{d^{cap}} \gets f^{cap}(p_{\vartheta_i}^{d,*})$}, from \eqref{eq:cap} \label{line:cap}\;
                {\small$EQT(\vartheta_i) \gets 1$\label{line:eqtgets1}}, \; {\small$EQP(\vartheta_i) \gets EQP(\vartheta_i)+p_{\vartheta_i}^{d^{cap}}$}\label{line:probincrease}\;
                 \label{line:pacomputed}
                    $X_\Sigma \gets \,X_\Sigma+ (N-1)~r_{\vartheta_i}~p_{\vartheta_i}^{d^{cap}}~E_{\vartheta_i}$
                }
            }\If{$|EQP- EQP_{old}| \leq tol$ \label{line:equilibrium}}{
                Exit\;
        }
    }
    \caption{Distributed algorithm for NE.}
    \label{algorithm}
\end{algorithm}
\end{small}

Further privacy concerns can be handled by encrypting the values of $EQT$ and $EQP$ at each iteration and appropriately authenticating users that will be able to decrypt only the entries of $EQT$ and $EQP$ that correspond to their type. However, if the first and second consumers to play are of the same type, then the second in row consumer may infer the type of the first one. To avoid this we should enforce that the second consumer type to play does not have the same energy profile as the first one. 
In the special case of $N=M$, broadcasting $EQT$ and $EQP$ is not needed; the computing consumer requires only the current value of $X_\Sigma$.

Finally, this algorithm schedules consumers' loads in daytime and nighttime intervals once, namely in the beginning of a daytime interval. However, in future work, we intend to study its repetition in a Model Predictive Control fashion over an intra-day time scale with time intervals of several hours, where at each repetition: (i) the consumers reconsider their daily energy demand profiles and exclude already served loads, (ii), the consumers reconsider their risk aversion degrees, and (iii) the forecast $\mathcal{RE}$ of the RES is updated. 

%% file: evaluation.tex
\vspace{-0.1in}
\section{Numerical Evaluations}
\label{sec:eval}
\vspace{-0.05in}
\subsection{Case Study Setup}
\vspace{-0.03in}
We consider a smart grid with $N=1000$ consumers, divided into $5$ distinct consumer types, with a maximum daytime energy demand $D^{Total} = 4250$ kWh. Table \ref{tab:residential} summarizes the consumer types parameters.
\begin{table}[t]
    \centering
    \begin{small}
    \begin{tabular}{|c||c|c|c|c|c|}
        \hline
        Type $\vartheta$ & 0 & 1 & 2 & 3 & 4 \\
        \hline 
        $E_\vartheta$ (kWh)& 2 & 3 & 5 & 10 & 15 \\
        \hline
        $r_\vartheta$ & 0.20 & 0.40 & 0.30 & 0.07 & 0.03 \\
        \hline
    \end{tabular}
    \caption{Game parameters for residential smart-grid.}
    \label{tab:residential}
    \end{small}
    \vspace{-0.2in}
\end{table}
The consumer type distribution and the daytime energy demand levels are selected to be consistent with European households \cite{enerdata}. Most households are moderately energy efficient (types $1$ and $2$), combined with many highly efficient households (type $0$) and few inefficient ones (types $3$ and $4$). Consumers of type $0$ are assumed to be risk-seeking ($\varepsilon_0=1$) and the risk-aversion degrees of all other types are determined by  \eqref{eq:relation_E_0_E_1_pa_ne_extra_demand}, but are close to $1$. We set the RES price as $c^{RES}=1$ \euro/kWh.

The proposed D-ESM with the PA policy is compared to a D-ESM with the ES policy for reference. Under ES, a so-called \textit{fair share} of RESs capacity is computed as

\vspace{-0.1in}
\begin{small}
\begin{align} \label{eq:fairshare}
 sh(\mathbf{p^{NE}}) 
& =\frac{\mathcal{RE}}{ N \sum_{ \vartheta \in \Theta} r_{\vartheta} ~ p_{\vartheta}^{d,NE}}.
\end{align}
\end{small}
\vspace{-0.1in}

\noindent 
Under ES, consumers of type $\vartheta \in \Theta$ that compete for $RES$ and have a daytime demand $E_{\vartheta} \leq sh(\mathbf{p^{NE}})$ are allocated their full daytime demand $E_{\vartheta} $, as well as an extra energy equal to $sh(\mathbf{p^{NE}})-E_{\vartheta}$ that will remain unused. 
On the contrary, the consumers of type $\vartheta \in \Theta$ that play the pure strategy $d$ and have a daytime demand $E_{\vartheta}>sh(\mathbf{p^{NE}})$ will be allocated the fair share and their remaining daytime energy demand $E_{\vartheta}-sh(\mathbf{p^{NE}})$ will be served by the highly priced peak-load generation. Therefore, the share of RESs received by a consumer $i$ of type $ \vartheta_i \in \Theta$ that plays the strategy $d$ is $rse^{ES}_{\vartheta_i}(\mathbf{p^{NE}}) = \min\left( E_{\vartheta_i}, sh(\mathbf{p^{NE}}\right))$. Note that this allocation policy may result in large inefficiencies due to unused RES capacity, even when the total aggregate demand for RESs $D (\mathbf{p^{NE}})$ is higher than $\mathcal{RE}$. Therefore, this allocation policy is solely used as a base-case comparison to the PA allocation. The definitions and/or analysis of the D-ESM and the C-ESM under the ES policy are provided in the Appendix C of \cite{arxiv_version}.

Finally, the convergence properties of the proposed decentralized algorithm under PA are studied for three capping systems, namely, (i) equal cap: $cap$ stays constant and equal to $0.1$; (ii) random cap: $cap$ is sampled from the uniform distribution $cap \sim U (0,1)$ (evaluated over multiple trials with varying values of $cap$); and (iii) no cap: equivalent to $cap=1$.

\vspace{-0.15in}
\subsection{Numerical Results}

\subsubsection{Social Cost and PoA under Varying Parameters}
\vspace{-0.07in}

The first set of numerical evaluations studies the proposed D-ESM under various tariff values, namely with $\beta = \{2 , 2.5 \}$ and $\gamma = 3$, as well as under varying available RES capacities $\mathcal{RE}$, ranging from $5\%$ to $125\%$ of $D^{Total}$.

\begin{figure}[ht!]
\vspace{-0.1in}
     \centering
     \subfigure[Social Cost (in eurocents). \label{fig:eval_1a}]{
         \includegraphics[width=0.23\textwidth]{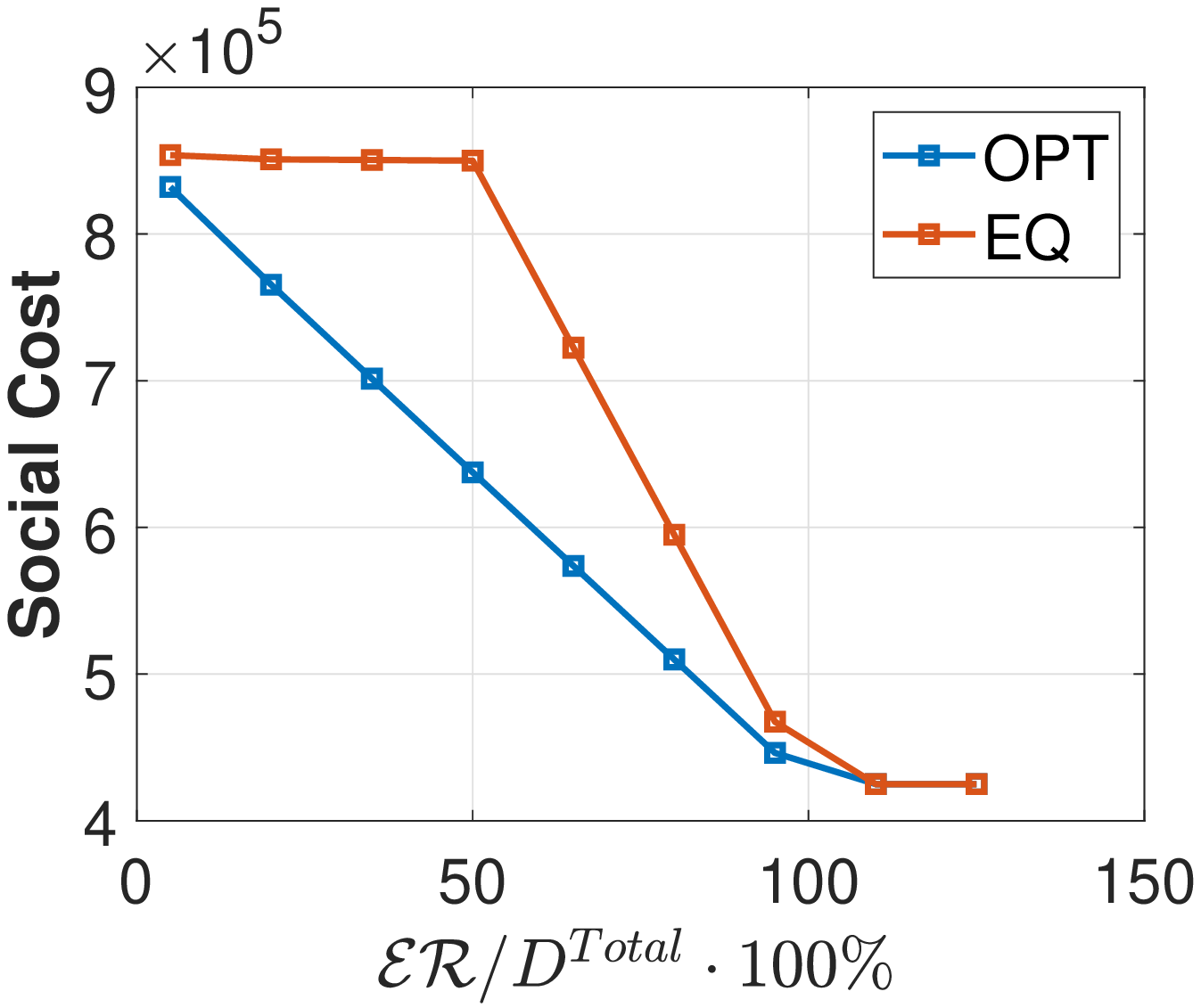}}
  \subfigure[PoA. \label{fig:eval_1b}]{
         \includegraphics[width=0.23\textwidth]{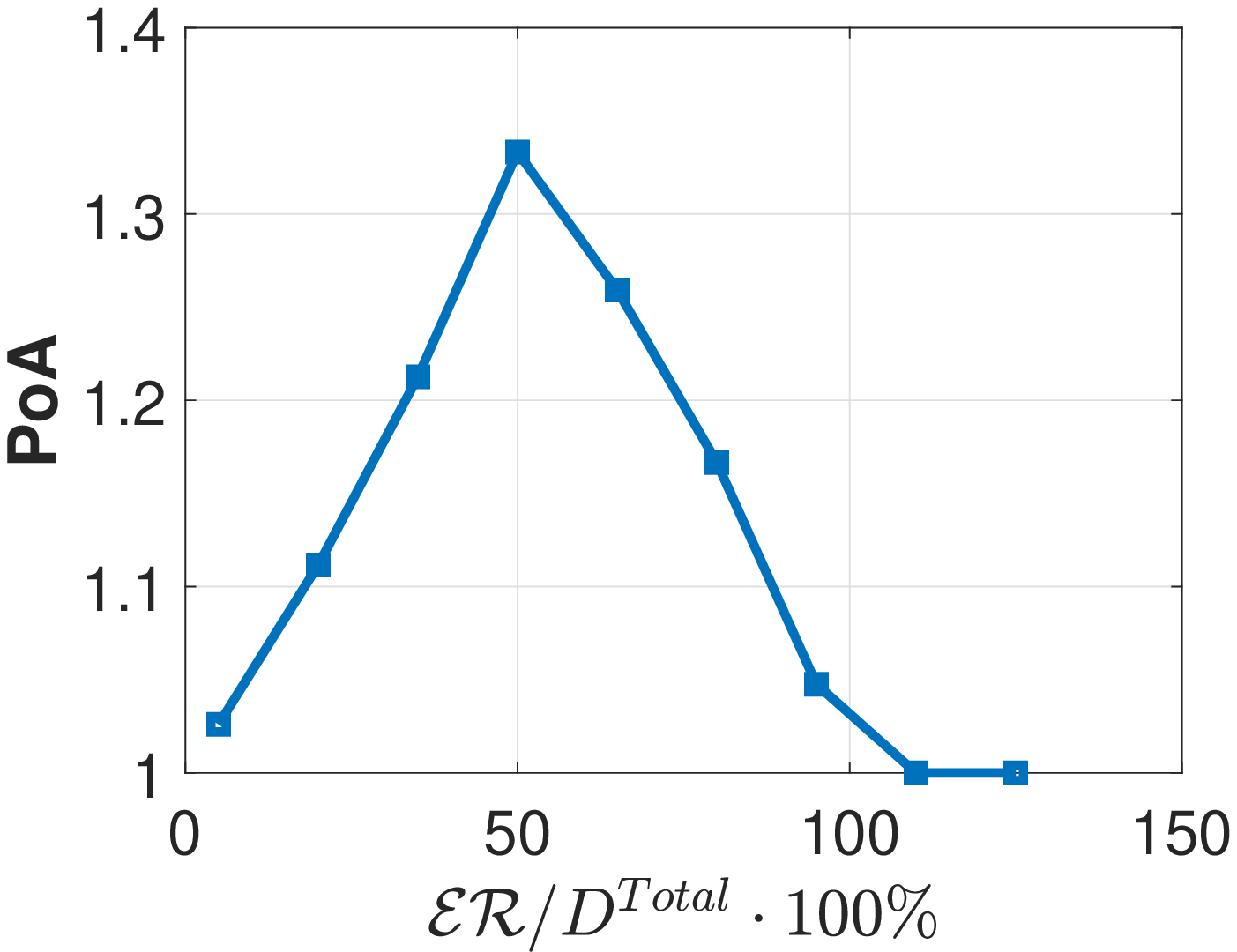}}
        \caption{Social cost and PoA under PA rule for residential grid with $\beta=2$.}
        \label{fig:eval_1}
        \vspace{-0.2in}
\end{figure}

As illustrated in Fig. \ref{fig:eval_1a}, the optimal social cost (given by Eq. \eqref{eq:social_cost_pa_extra_demand_2}) as derived by C-ESM, denoted by OPT, decreases linearly with $\mathcal{RE}$. Indeed, since all risk-aversion degrees are equal or close to $1$, the cost function can be approximated as {\small $ \mathcal{RE} (1-\gamma) c^{RES} +N \sum_{\vartheta \in \Sigma_2} \left[ r_{\vartheta} E_{\vartheta} \left(\gamma -  \beta \right) p^{d,*}_{\vartheta} \right] c^{RES}+ D^{Total} \beta c^{RES} \approx \mathcal{RE}(1-\beta) c^{RES}+ D^{Total} \beta c^{RES}$}, which is constant with respect to the competing probabilities and linearly decreasing with $\mathcal{RE}$. Note that since $\gamma=3$, $\beta=2$ and all risk aversion degrees are close to $1$, all consumers belong in the set $\Sigma_2$. Furthermore, we have observed that the minimization by C-ESM results in "big players" competing for RESs (i.e., playing the strategy $d$) at the expense of smaller ones. It is indeed observed that consumers with lower daytime energy demand play the strategy $d$ with non-zero probability only if there is remaining RES capacity when all consumers with higher daytime energy demand compete for RES with probability 1. This is aligned with the theoretical solution of the C-ESM in Section \ref{sec:centralsol}, since according to Remark 2, the larger the daytime energy demand of the player is the lower her risk aversion degree should be. Thus, larger players are prioritized in getting the highest probabilities values for competing for RES also according to the theoretical analysis.   

On the other hand, as seen in Fig. \ref{fig:eval_1a}, for the D-ESM, the social cost is almost constant with the initial increase in the RES capacity due to the fact that consumers tend to over-compete for RES (i.e., play more often the strategy $d$) as can be observed in the obtained values of the competing probabilities. However, for $\mathcal{RE}\in [0.5 D^{Total}, D^{Total}]$, the social cost decreases when $\mathcal{RE}$ increases, because there exists less excess demand for RES and thus the amount of required highly priced daytime non-RES energy is reduced.

As illustrated in Fig. \ref{fig:eval_1b}, the PoA values are rather small for all values of $\mathcal{RE}$. The PoA peaks for $\mathcal{RE} \approx 0.5 \cdot D^{Total}$, which is the point at which the social cost for the decentralized mechanism begins decreasing. This graph can provide valuable insights into how much RES capacity should be installed to increase the efficiency of the D-ESM. We can identify two zones of high efficiency, namely for low and high RES capacity. In the first zone, this is due to the small gains in cost offered by low RES capacity in both the centralized and the decentralized mechanisms. In the second zone, the NE solution has almost converged to the optimal solution and thus social costs are optimal. 

\begin{figure}[t] 
\vspace{-0.1in}
     \centering
     \subfigure[Social Cost (in eurocents). \label{fig:eval_2a}]{
         \includegraphics[width=0.23\textwidth]{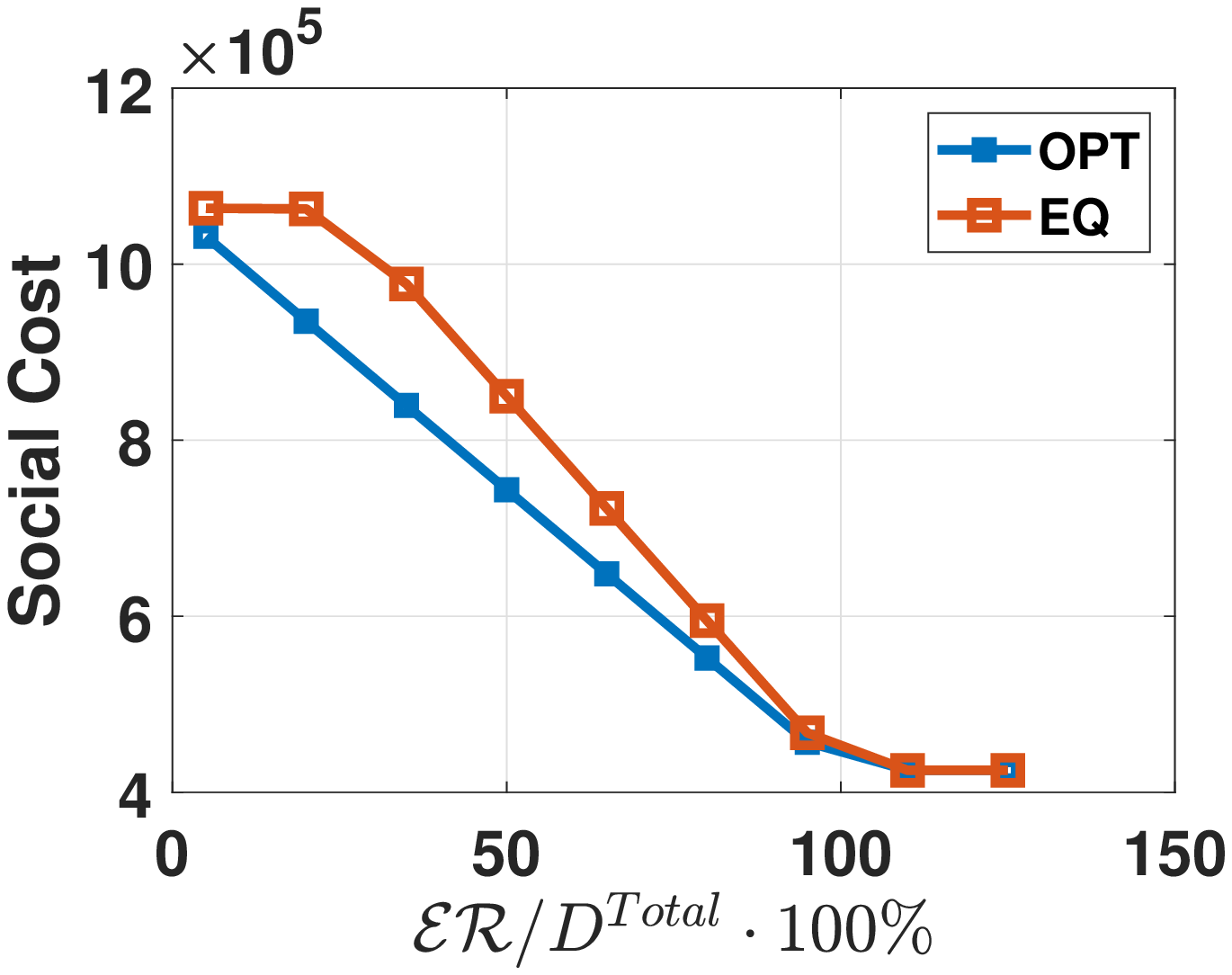}}
  \subfigure[PoA. \label{fig:eval_2b}]{
         \includegraphics[width=0.23\textwidth]{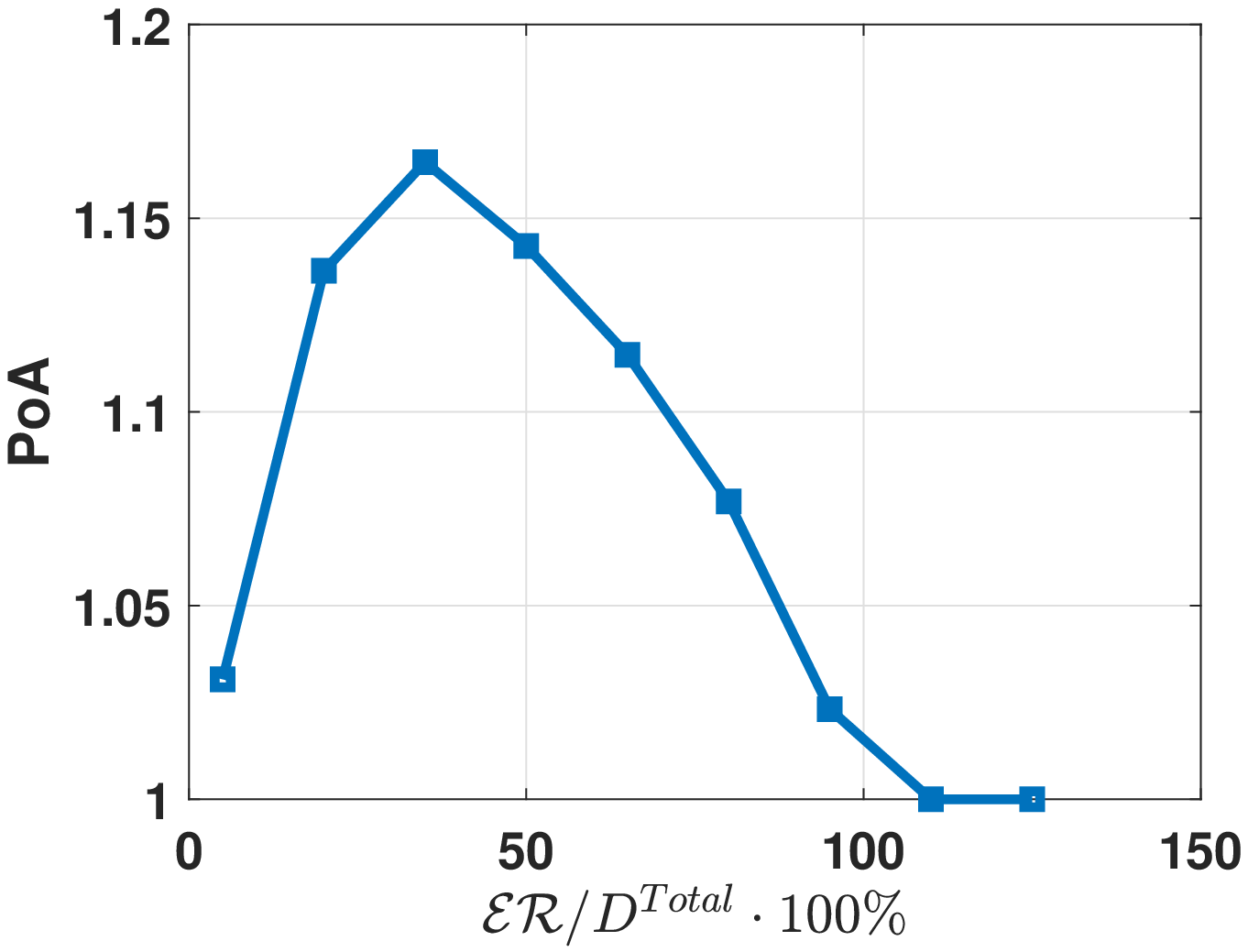}}
        \caption{Social cost and PoA under PA rule for residential grid with $\beta=2.5$.}\vspace{-0.2in}
        \label{fig:eval_2}
\end{figure}

In addition, the value of $\mathcal{RE}$ at which the PoA reaches its peak (most inefficient outcome) depends on the system model parameters and most importantly on the price parameters $\beta$ and $\gamma$. 
In particular, from Fig. \ref{fig:eval_2a} we observe that the cost values of both C-ESM and D-ESM are higher for $\beta=2.5$, compared to $\beta=2$ (Fig. \ref{fig:eval_1b}), because setting  $\beta=2.5$ results in higher night-time costs.
However, for $\beta=2.5$, the cost curve of D-ESM starts decreasing at lower values of available RES capacity, namely
at $\mathcal{RE} = 20 \% \cdot D^{Total}$. Moreover, as seen in Fig. \ref{fig:eval_2b}, the PoA attains significantly lower values for higher $\beta$ and peaks at
around $1.16$. Therefore, when the nighttime cost increases, D-ESM behaves closer to the optimal solution. More results on how the tariff values' changes (via the parameters $\gamma$ and $\beta$) affect the NE can be found in \cite{stai2022}.

\begin{figure}[t] 
     \centering
         \includegraphics[width=0.32\textwidth]{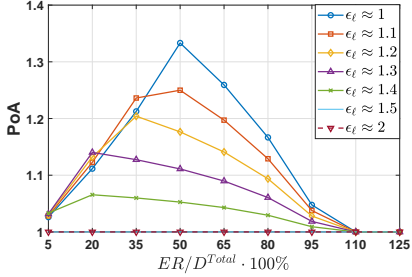}
        \caption{PoA vs (inverse) risk aversion degree.}
        \label{fig:eval_poarisk}
        \vspace{-0.2in}
\end{figure}

Finally, in Fig. \ref{fig:eval_poarisk}, the PoA is compared for different values of risk aversion of the energy community with $\beta=2$.  In particular, the inverse risk aversion degree of consumers of type $0$ are set to values between $\varepsilon_0=1$ and $\varepsilon_0=2$ (as indicated in the legend) and the risk-aversion degrees of all other consumer types are determined by  \eqref{eq:relation_E_0_E_1_pa_ne_extra_demand}. It turns out that all inverse risk-aversion degrees are either equal or very close to $\varepsilon_0$, and, thus, the consumers in the energy community have all approximately the same risk aversion. It can be observed that as consumers become less risk seeking (i.e., $\varepsilon_{\vartheta}$ increases and thus $\mu_{\vartheta}$ decreases), the PoA values decrease for all $\mathcal{RE} /D^{Total}$ ratios exceeding $50\%$ in this plot. Thus, our proposed distributed scheme reveals that the achieved social cost of a less risk seeking community moves closer to the optimal for all possible NE and in particular, for $\varepsilon_{\vartheta}\geq 1.5$ (or for $\mu_{\vartheta}\leq 0.67$) the PoA values are optimal (i.e., equal to $1$) for all values of $\mathcal{RE} /D^{Total}$. As a conclusion, under conditions such as those associated with Fig. \ref{fig:eval_poarisk}, less risk seeking behavior by the community can yield NE inducing a social cost arbitrarily close to the optimal (PoA be reduced to as low as 1). Notice from Fig. \ref{fig:eval_poarisk} that a deviation of the social cost of about $33\%$ from the optimal social cost ($PoA=1.33$ for $\varepsilon_{\vartheta}=1$ and $\mathcal{RE} /D^{Total}$ ratio of $50\%$) can be entirely eliminated by adopting a less risk seeking behavior ( $\varepsilon_{\vartheta}\geq 1.5$).

\subsubsection{Comparison to the ES Policy}
\label{sec:comptoES}
\begin{figure}[t] 
     \centering
     \subfigure[Social Cost (in eurocents). \label{fig:eval_3a}]{
         \includegraphics[width=0.23\textwidth]{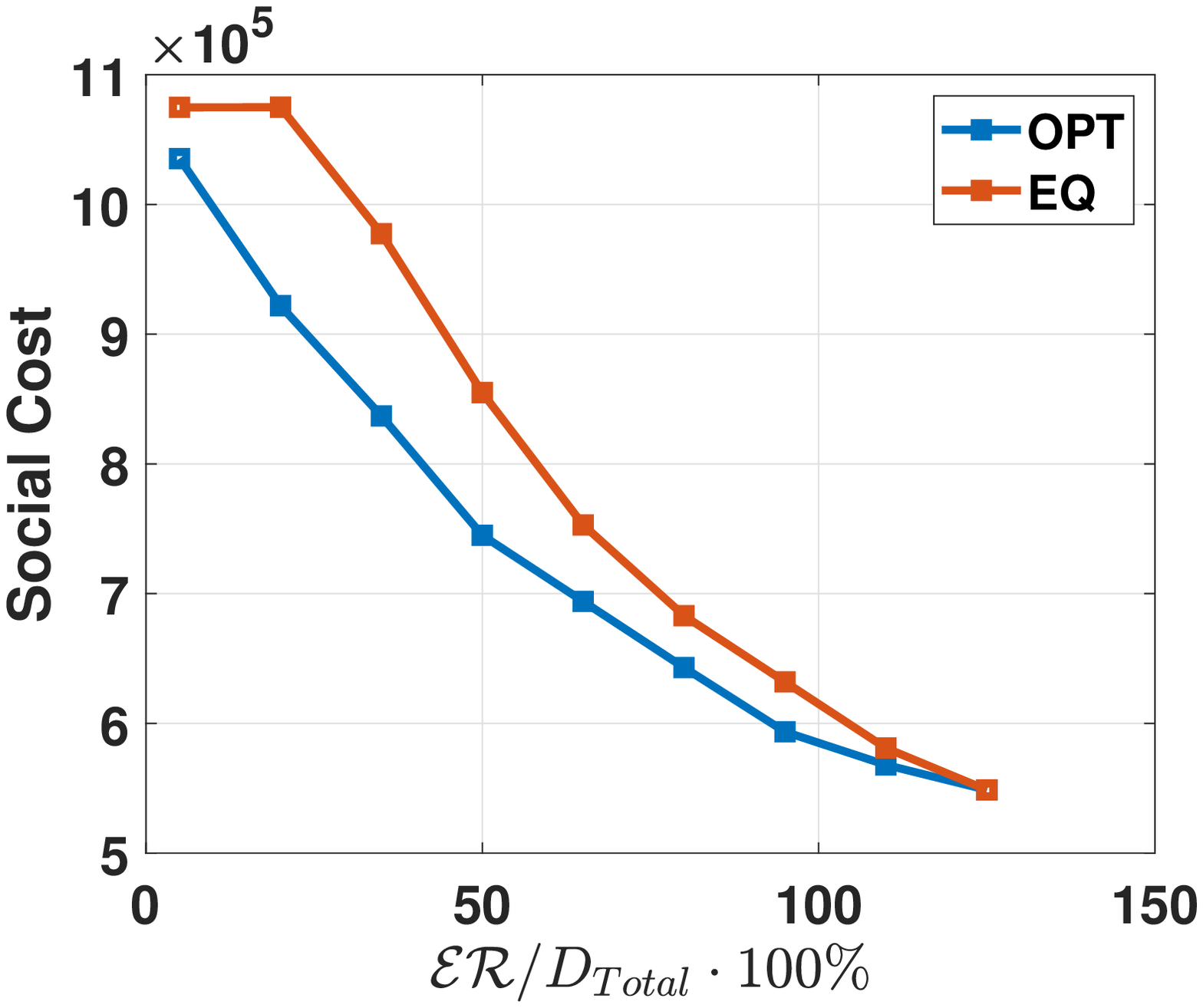}}
  \subfigure[PoA. \label{fig:eval_3b}]{
         \includegraphics[width=0.23\textwidth]{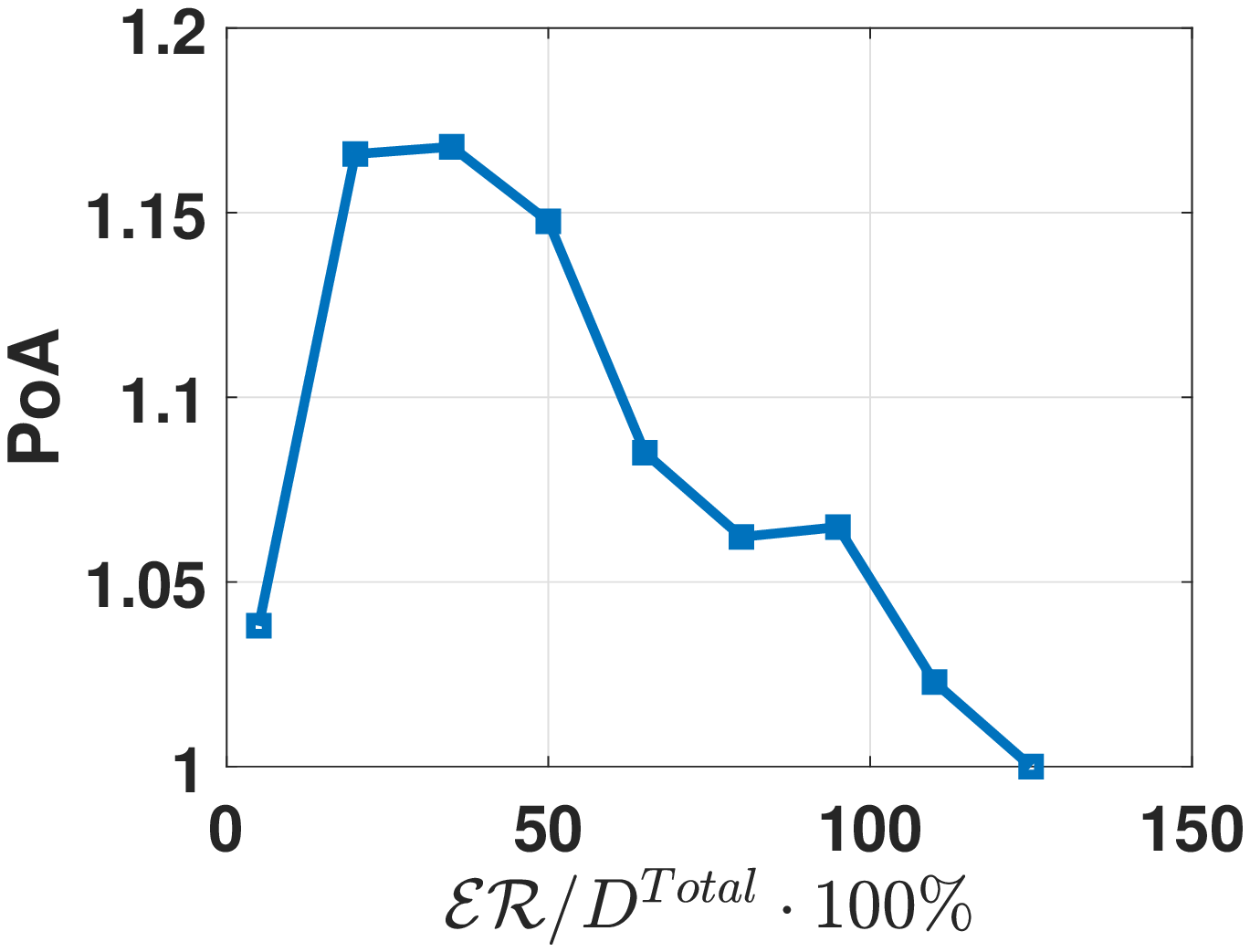}}
        \caption{Social cost and PoA under ES for residential smart grid with $\beta=2$.}
        \label{fig:eval_3}
        \vspace{-0.2in}
\end{figure}

As observed in Fig. \ref{fig:eval_3a}, both C-ESM and D-ESM yield higher social costs under the ES than under the PA policy for all values of RES capacity.

This is due to i) the unused RES capacity by consumers' types whose demand for RES is lower than the fair share; and ii) the resulting increased daytime non-RES energy needed to cover the unsatisfied demand of consumers' types whose demand for RES is higher than the fair share.
In addition, with ES, even for $\mathcal{RE}=125 \% \cdot \mathcal{RE}_{max}$ and even for the centralized mechanism the competing probabilities may not be all equal to $1$. The mechanism may reduce the competing probabilities of smaller players in order to increase the RES utilization. The social cost of both D-ESM and C-ESM decrease with increasing $\mathcal{RE}$, but not linearly contrary to the PA policy, due to the non-linearity of the cost functions with respect to $\mathcal{RE}$ under the ES policy. Moreover, we observe that the social cost of D-ESM under ES follows a similar trend as under PA, namely, it is constant for small values of $\mathcal{RE}$ and then starts to decrease. This shows that, similarly to the PA rule, consumers tend to over-compete for RES under the ES policy, especially for lower values of the RES capacity.

Additionally, as seen in Fig. \ref{fig:eval_3b}, the ES policy achieves lower PoA than the PA policy for most values of the RES capacity. However, the D-ESM under ES achieves $100\%$ efficiency only when the RES capacity reaches $\mathcal{RE}=125 \% \cdot D^{Total}$, whereas, for the PA policy, the PoA is equal to $1$ for lower values of RES capacity $\mathcal{RE} \geq 110\% \cdot D^{Total}$. Hence, using the ES policy may be more expensive in case that $100\%$ efficiency of the D-ESM is required for which it requires more RES capacity than PA. Furthermore, due to the non-linearity of the social cost function with $\mathcal{RE}$, the PoA curve does not decrease monotonously after the initial peak.


\subsubsection{Evaluation of Distributed Algorithm}

Here, we evaluate the performance and convergence of Algorithm \ref{algorithm}. For easier visualization, we have implemented the algorithm in a smart grid with $N=500$ consumers divided into two consumer types, using the following parameter values: $E_0=100$ kWh, $E_1=200$ kWh, $D^{Total}=65000$ kWh, $r_0=0.7$, $r_1=0.3$, $\varepsilon_0=1$, $\varepsilon_1=1.004$, $c^{RES}=100$ \euro/kWh, $\beta=2$, $\gamma=4$, and $\mathcal{RE}=25\% \cdot D^{Total}=16250$ kWh.

Table \ref{tab:algo_sc_PA} summarizes the evaluation results on the social cost, the aggregate daytime energy demand (Eq. \eqref{eq:demand1_2c}) and the PoA for the optimal centralized solution as well as for the solution of the distributed algorithm for the three capping systems.

\begin{table}[b] \vspace{-0.15in}
\begin{small}
    \resizebox{\linewidth}{!}{
        \begin{centering}
    \begin{tabular}{|c||c|c|c|c|}
        \hline
        & \textbf{Social Cost} ($10^6$) & \textbf{Demand} & \textbf{PoA} & \textbf{Number of steps} \\
        \hline 
        Centralized & 11.37 & 16250 & 1 & - \\
        \hline
        Equal cap & 13.00 & 24321 & 1.14 & 18  \\
        \hline
        Random cap & 12.99 & 24286 & 1.14 & 17-27 \\
        \hline
        No cap & 13.01 & 24324 & 1.14 &  14 \\
        \hline
    \end{tabular}
    \end{centering}
    }
    \caption{Social cost, demand, PoA, and number of iterations until convergence under the PA rule for centralized and distributed algorithmic solutions.}
    \label{tab:algo_sc_PA}
    \vspace{-0.15in}
    \end{small}
\end{table}

All three capping methods lead to similar social cost and PoA values. Thus, the choice of capping method mostly influences the competing probabilities to introduce an additional fairness level for sharing the RES capacity among the consumer types, without affecting the social cost. To clarify, the fairness level introduced by the capping system is with respect to the mixed strategies level due to the fact that the order that consumers play has an influence; whereas the fairness of the allocation policy is with respect to the assignment of the available RES to those that finally compete for RES. Furthermore, Table \ref{tab:algo_sc_PA} highlights that if we do not apply a capping scheme the algorithm converges the fastest \footnote{The tolerance is set to $tol=10^{-4}$.} at the expense of fairness. This is because we do not restrict the rate at which the solution reaches a NE. Introducing a constant capping system slightly deteriorates convergence, but it stays within the same order of magnitude. Lastly, the random capping system provides no control over the convergence speed, and we observe a large variance in the required number of steps (outer loops) to convergence. Note that lower $cap$ values increase the required number of steps for convergence. Most importantly, for all three capping systems, we observe that the number of steps until convergence is much lower than the number of players ($N=500$), which showcases the efficiency of the algorithm.

\begin{figure}[t] 
     \centering
     \includegraphics[width=0.27\textwidth]{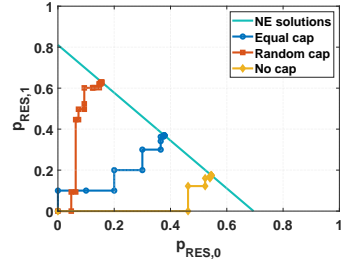}
     \caption{Decentralized algorithmic solutions for different capping systems.}\label{fig:eval_4a}
    \vspace{-0.2in}
\end{figure}

Figure \ref{fig:eval_4a} illustrates the solution paths given by the distributed algorithm for all three capping systems. It can be observed that all solution paths converge to a theoretically proven NE, represented by the blue line. For the constant cap ($cap=0.1$), the solution path oscillates around the $45^{\circ}$ line. Therefore, the achieved NE solution consists of similar competing probability values for both consumer types. Lower constant $cap$ values increase fairness among consumer types, and greatly dampen any bias towards any type. If the random cap method is implemented, the solution path is naturally random. Lastly, with the no cap system, the consumer type that plays first gains a considerable advantage.

%% file: conclusion.tex
\vspace{-0.1in}
\section{Conclusions}\vspace{-0.05in}
\label{sec:conclusions}

In this paper, we analyze the uncoordinated decisions of self-interested risk-aware consumers participating in an energy sharing community and a decentralized ESM through a non-cooperative game-theoretic framework. We prove the existence of dominant solutions and/or NE for different energy tariff values and renewable energy allocation policies, as well as under various levels of consumers' risk-aversion and energy demand. 
For low and medium values of RESs production, consumers are shown to over-compete for RESs compared to the optimal solution giving rise to higher cost values. However, the incorporation of consumers’ attitude toward risk in the model considered in this work has revealed that the PoA peaks can be reduced and even alleviated as the energy community becomes more risk conservative. Moreover, the PA policy outperforms ES in terms of social cost. Finally, choosing a fair NE among all possible ones is also studied using a distributed algorithm for choosing consumers' actions.

From a methodological point of view, a natural direction for further investigation is to account for more complex behavioral human-driven models of consumers' decision-making (\eg \cite{McKP95}). 
Also, we could incorporate and compare various tariff schemes, which account both for competition across multiple energy sources and time steps. Finally, the authors in \cite{rodriguez2021value} showed that information has a major impact on the efficiency loss in decentralized DRPs. Thus, future work will analyze consumer competition in an energy community under more realistic assumptions of imperfect information.

%% file: appendix_new.tex
\newpage

\section{Proofs for Case $2$ of the D-ESM}
\label{sec:proofsESSG}
\vspace{-0.05in}
For the consumers in $\Sigma_1$, we need to show that  $\upsilon^d_{\vartheta}(\mathbf{p})<\upsilon^n_{ \vartheta}(\mathbf{p})$, $\forall \mathbf{p}$ and $\forall \vartheta \in \Sigma_1$. Assume a consumer type $\vartheta\in \Sigma_{1}$ and that her allocated RES energy is $E'$. Then, we have that $\upsilon^d_{ \vartheta}(\mathbf{p})= E'  \cdot c^{RES}+(E_{\vartheta}-E')\cdot \gamma \cdot  c^{RES}$ and 
$\upsilon^n_{ \vartheta}(\mathbf{p})= \varepsilon_\vartheta \cdot E_{\vartheta} \cdot \beta \cdot c^{RES}$. The inequality $\upsilon^d_{\vartheta}(\mathbf{p})<\upsilon^n_{ \vartheta}(\mathbf{p})$ is then equivalent to $ E'  (1-\gamma) \cdot c^{RES} <  E_\vartheta \cdot (\varepsilon_\vartheta \cdot \beta -\gamma) \cdot c^{RES}$, which is true by assumption, since $(1-\gamma)<0$ and $(\varepsilon_\vartheta \cdot \beta -\gamma)>0$.

  Next, for the consumers in $\Sigma_{2,1}$, we need to show that $\upsilon^d_{ \vartheta}(\mathbf{p})>\upsilon^n_{ \vartheta}(\mathbf{p})$, $\forall \mathbf{p}$ and $\forall \vartheta\in \Sigma_{2,1}$. Assume a consumer type $\vartheta \in \Sigma_{2,1}$ and that her allocated RES energy is $E'$. Then, the inequality $\upsilon^d_{\vartheta}(\mathbf{p})>\upsilon^n_{ \vartheta}(\mathbf{p})$ is equivalent to the inequality $E_\vartheta >E' \frac{(\gamma-1)}{(\gamma-\varepsilon_\vartheta\beta)}$, which is true by assumption, since $E'<\mathcal{RE}$.

Now, we prove the condition of existence of a mixed strategies NE for the consumers in $\Sigma_{2,2}$. Recall that in the ESG under the PA policy, a mixed strategy NE, $\mathbf{p^{NE}}$, among consumers in $\Sigma_{2,2}$ exists under the condition
\vspace{-0.05in}

\small
\begin{equation}\label{eq:condition_PA_NE_2}
res_{\vartheta}^{PA}(\mathbf{p}^{NE}) =res_{\vartheta}^{NE}(\mathbf{p}^{NE}), \forall \vartheta \in \Sigma_{2,2}. \end{equation}
\normalsize

To derive condition \eqref{eq:relation_E_0_E_1_pa_ne_extra_demand} we re-write \eqref{eq:condition_PA_NE} first with assuming that a consumer $i$ of type $\vartheta_i \in \Sigma_{2,2}$ plays the pure strategy $A_i=d$ (in \eqref{eq:probrelation1}) and second with assuming that a consumer $j$ with type $\vartheta_j \in \Sigma_{2,2} \setminus \{\vartheta_i\}$ plays the pure strategy $A_j=d$ (in \eqref{eq:probrelation2}):

\vspace{-0.1in}
\begin{small}
\begin{align}
&  \mathcal{RE}\frac{(\gamma-1)}{(\gamma-\varepsilon_{\vartheta_i}\beta)}-E_{\vartheta_i}= D^{Total}_{\Sigma_1}+\sum_{ {\vartheta'}\in \Sigma_{2,2}} r_{\vartheta'}~ (N-1)~E_{\vartheta'}~p^{d,NE}_{\vartheta'},
    \label{eq:probrelation1}\\
  &  \mathcal{RE}\frac{(\gamma-1)}{(\gamma-\varepsilon_{\vartheta_j}\beta)}-E_{\vartheta_j}=D^{Total}_{\Sigma_1}+  \sum_{ {\vartheta'}\in \Sigma_{2,2}} r_{\vartheta'} ~(N-1)~E_{\vartheta'}~ p^{d,NE}_{\vartheta'}.
    \label{eq:probrelation2}  
\end{align}
\end{small}

Note that to derive (\ref{eq:probrelation1}) we consider that if a consumer $i$ in $\Sigma_{2,2}$ of type $\vartheta_i$ plays  the pure strategy $A_i=d$, then, the aggregate expected daytime energy of the consumers in $\Sigma_{2,2}$, $D_{\Sigma_{2,2}}(\mathbf{p^{NE}})$ can be expressed as $E_{\vartheta_i}+ \sum_{ {\vartheta'}\in \Sigma_{2,2}} r_{\vartheta'}~ (N-1)~E_{\vartheta'}~p^{d,NE}_{\vartheta'}$ for a large number of consumers and similarly also for (\ref{eq:probrelation2}). Then, since the right-hand sides of \eqref{eq:probrelation1}-\eqref{eq:probrelation2} are equal, the left-hand sides will be also equal and \eqref{eq:relation_E_0_E_1_pa_ne_extra_demand} derives. 

To derive the probability bounds, we re-write \eqref{eq:condition_PA_NE} assuming that all consumers of the same type play the same mixed strategy, i.e., 

\vspace{-0.1in}
\begin{small}
\begin{align}
&  \mathcal{RE}\frac{(\gamma-1)}{(\gamma-\varepsilon_{\vartheta_i}\beta)}= D^{Total}_{\Sigma_1}+N\sum_{ {\vartheta'}\in \Sigma_{2,2}} r_{\vartheta'}~ E_{\vartheta'}~p^{d,NE}_{\vartheta'}.
    \label{eq:probrelation3}  
\end{align}
\end{small}

The minimum bound on the probability for competing for RESs, $p_{\vartheta}^{\min}$, derives by setting in (\ref{eq:probrelation3}) $p^{d,NE}_{\tilde{\vartheta}}=1$, $\forall \tilde{\vartheta}\in \Sigma_{2,2}$ with $\tilde{\vartheta}\neq \vartheta=\vartheta_i$. Similarly, the maximum bound on the probability for competing for RESs, $p_{\vartheta}^{\max}$, derives by setting in (\ref{eq:probrelation3}) $p^{d,NE}_{\tilde{\vartheta}}=0$, $\forall \tilde{\vartheta}\in \Sigma_{2,2}$ with $\tilde{\vartheta}\neq \vartheta=\vartheta_i$. 

Finally, the expression for the aggregate expected daytime energy demand given in \eqref{eq:demand1_2c} is constructed as follows. First we can write that 
\begin{align}
 D^{d,NE} = D^{Total}_{\Sigma_1} +N\sum_{ {\vartheta'}\in \Sigma_{2,2}} r_{\vartheta'}~E_{\vartheta'}~p^{d,NE}_{\vartheta'}. \label{eq:totdemand}
 \end{align}

Second, by multiplying \eqref{eq:probrelation1} with $\frac{N}{N-1}$, we obtain:

\begin{small}
\begin{align}
&  N\sum_{ {\vartheta'}\in \Sigma_{2,2}} r_{\vartheta'}~E_{\vartheta'}~p^{d,NE}_{\vartheta'}=\frac{N}{N-1}\left[\frac{\mathcal{RE}(\gamma-1)}{(\gamma-\varepsilon_{\vartheta_i}\beta)}-E_{\vartheta_i}-D^{Total}_{\Sigma_1}\right].
    \label{eq:probrelation3}
\end{align}
\end{small}

Third, by replacing \eqref{eq:probrelation3} in \eqref{eq:totdemand} we obtain \eqref{eq:demand1_2c}, where the $\min\{.\}, ~\max\{.\}$ operators account for the case that the initially obtained probability values by  \eqref{eq:probrelation1} do not lie in the range $[0,1]$ and should be set to the values $1$ or $0$, correspondingly. 

\section{Proofs for Case $2$ of the C-ESM}
\label{appendix:dual}

In this case, it is optimal for the C-ESM to schedule loads during the day so that the total RES capacity is fully utilized, i.e., the expected aggregate daytime energy demand is greater than or equal to the RES capacity:

 \vspace{-0.1in} 
 \small
\begin{align}
N \sum_{{\vartheta} \in \Theta}r_{{\vartheta}} ~E_{{\vartheta}}~p^{d}_{{\vartheta}} \geq \mathcal{RE}.
\label{eq:optimal}
\end{align}
\normalsize 
\vspace{-0.1in}

\noindent Therefore, the social cost reduces to:

 \vspace{-0.1in} 
 \begin{small}
\begin{align}
C(\mathbf{p}) &=  \mathcal{RE} \cdot c^{RES}   + \left[N \sum_{{\vartheta} \in \Theta} r_{\vartheta} ~p_{{\vartheta}}^{d}~  E_{\vartheta} - \mathcal{RE}\right]  \gamma \cdot c^{RES} \nonumber \\
& + N \left[ \sum_{{\vartheta} \in \Theta} r_{\vartheta } \left(1-p^{d}_{{\vartheta}}\right)\varepsilon_{{\vartheta} }~E_{{\vartheta}}\right] \beta \cdot c^{RES},
 \label{eq:social_cost_pa_extra_demand_2}
 \end{align}
\end{small} \vspace{-0.1in}

\noindent and the C-ESM optimization problem \eqref{eq:social_cost_x_opt} is equivalent to minimizing $ N \sum_{\vartheta \in \Theta} \left[ r_{\vartheta} E_{\vartheta} \left(\gamma - \varepsilon_{\vartheta} \beta \right) p^{d}_{\vartheta} \right] c^{RES}$, subject to constraints \eqref{eq:opt_2.1}-\eqref{eq:opt_3.2} and \eqref{eq:optimal}. Below, we derive closed-form expressions of the solutions of this linear optimization problem.

We define two complementary subsets of consumer types, depending on their risk aversion degrees: $\Sigma_1 = \Bigl\{\vartheta \in \Theta : \varepsilon_{\vartheta} \geq \gamma/\beta \Bigr\} \subset \Theta$, and $\Sigma_2 = \Bigl\{ \vartheta \in \Theta :1\leq \varepsilon_{\vartheta} < \gamma/\beta \Bigr\} \subset \Theta$.

For all consumers whose type $\vartheta \in \Sigma_1$, it is optimal for the C-ESM to schedule them during daytime, such that $p^{d,*}_{\vartheta}=1 $. Therefore, the optimal schedule for the remaining consumers whose type $\vartheta \in \Sigma_2$ can be found by solving the following linear optimization problem:

 \begin{small}
 \begin{subequations} \label{eq:social_cost_x_opt_2}
\begin{alignat}{2}
& \min_{\mathbf{p}} \ &&  N \sum_{\vartheta \in \Sigma_2} \left[ r_{\vartheta} ~E_{\vartheta} \left(\gamma - \varepsilon_{\vartheta} \beta \right) p^{d}_{\vartheta} \right] c^{RES} \label{eq:opt_S2_1} \\
 & \text{s.t. } && \eqref{eq:opt_2.1}-\eqref{eq:opt_3.2} \label{eq:opt_S2_2}\\
 & \quad && N \sum_{{\vartheta} \in \Sigma_2}r_{{\vartheta}} ~E_{{\vartheta}}~p^{d}_{{\vartheta}} \geq \left( \mathcal{RE} - N \sum_{{\vartheta} \in \Sigma_1}r_{{\vartheta}} E_{{\vartheta}}\right). \label{eq:opt_S2_3} 
 \end{alignat}
 \end{subequations}
\end{small} 

\noindent And the dual function of this optimization problem is 

\begin{footnotesize}
 \begin{align} \label{eq:social_cost_x_opt_2_dual}
 \max_{\lambda \geq 0}\min_{\mathbf{p}} \quad & N \sum_{\vartheta \in \Sigma_2} \left[ r_{\vartheta} E_{\vartheta} \left(\gamma - \varepsilon_{\vartheta} \beta \right) p^{d}_{\vartheta} \right] c^{RES} \nonumber \\&-\lambda\left( N \sum_{{\vartheta} \in \Sigma_2}r_{{\vartheta}} E_{{\vartheta}}p^{d}_{{\vartheta}} - \left( \mathcal{RE} - N \sum_{{\vartheta} \in \Sigma_1}r_{{\vartheta}} E_{{\vartheta}}\right)\right),
 \end{align}
\end{footnotesize} \vspace{-0.1in}

\hspace{-0.2in} subject to \eqref{eq:opt_S2_2}, where $\lambda$ represents the dual variable associated with \eqref{eq:opt_S2_3} and let $\lambda^*$ represent its optimal value.

It results that:\\
$\bullet$ for all $\vartheta \in \Sigma_{2}$ where $1 \leq \varepsilon_\vartheta < \dfrac{\gamma c^{RES} - \lambda^*}{\beta c^{RES}}$, $p^{d,*}_{\vartheta}=0$,\\
$\bullet$ for all $\vartheta \in \Sigma_2$ where $ \varepsilon_\vartheta = \dfrac{\gamma c^{RES} - \lambda^*}{\beta c^{RES}}$, $0<p^{d,*}_{\vartheta}<1$,\\
$\bullet$  for all $\vartheta \in \Sigma_2$ where $ \dfrac{\gamma c^{RES} - \lambda^*}{\beta c^{RES}} < \varepsilon_\vartheta <\dfrac{\gamma}{\beta}$, $p^{d,*}_{\vartheta}=1$.

This means that the consumer types are fully dispatched during the day in the order of increasing risk aversion degree (or decreasing $\varepsilon_\vartheta$), until constraint \eqref{eq:opt_S2_3} is satisfied.

\section{Analysis For the ES Allocation Policy}
\subsection{Decentralized Energy Sharing Mechanism Under ES}
The analysis and proofs of this section follow similar lines as the analysis and proofs for the PA policy. Most proofs are however omitted for brevity.

In the ESG with the ES policy, a mixed-strategy NE exists under the condition:

 \small
\begin{equation}\label{eq:condition_ES_NE}
rse^{ES}_{\vartheta_i}(\mathbf{p^{NE}}) =res_{\vartheta}^{NE}(\mathbf{p}^{NE}), ~\forall \vartheta \in \Theta. \end{equation}
\normalsize 
\vspace{-0.1in}  

\noindent 
Let us distinguish the following cases:

\subsubsection*{\textbf{Case $1$: $\bm{\mathcal{RE}}$ exceeds $\bm{D^{Total}}$}}

As consumers have knowledge of $\mathcal{RE}$ and $D^{Total}$, it is straightforward to show that the dominant-strategy for all consumers is to schedule their daily flexible loads during daytime. As a result, the competing probabilities that lead to equilibrium states are equal to $p_{\vartheta}^{d,NE} = 1$ for all consumer types $\vartheta \in \Theta$.

\subsubsection*{\textbf{Case $2$: $\bm{\mathcal{RE}}$ is lower than $\bm{D^{Total}}$}} 

In this case, the strategies of the consumers depend on their respective risk aversion degrees and the TOU tariffs. We define two complementary subsets of consumer types, depending on their risk aversion degrees: $\Sigma_1 = \Bigl\{\vartheta \in \Theta : \varepsilon_{\vartheta} \geq \gamma/\beta \Bigr\} \subset \Theta$, and $\Sigma_2 = \Bigl\{ \vartheta \in \Theta :1\leq \varepsilon_{\vartheta} < \gamma/\beta \Bigr\} \subset \Theta$.

Firstly, the dominant strategy for all consumers $i$ whose type $\vartheta_i$ is in the set $\Sigma_1$ is to schedule their daily flexible loads during daytime, i.e., to play the pure strategy $A_i = d$ with probability $p_{\vartheta_i}^{d,NE} = 1$. Their expected aggregate daytime energy demand is then $D^{Total}_{\Sigma_1}=N\sum_{\theta \in \Sigma_1}r_{\theta} E_{\theta}$.

Secondly, the strategies of the consumers $i$ whose type $\vartheta_i$ is in the set $\Sigma_2$ depends on their daily flexible loads and risk-aversion degrees. Therefore, we define two distinct subsets of consumer types in $\Sigma_2$: $\Sigma_{2,1} = \left\{ \vartheta \in \Sigma_2 : E_{\vartheta} > \mathcal{RE}\frac{(\gamma-1)}{(\gamma-\varepsilon_{\vartheta}\beta)} \right\}$ and $\Sigma_{2,2} = \left\{ \vartheta \in \Sigma_2 : E_{\vartheta} \leq \mathcal{RE}\frac{(\gamma-1)}{(\gamma-\varepsilon_{\vartheta}\beta)}\right\}$.


For consumers $i$ whose type $\vartheta_i$ is in the set $\Sigma_{2,1}$, the dominant strategy is to schedule their daily flexible loads during nighttime, i.e., to play the pure strategy $A_i=n$ with probability $p^{n,NE}_{\vartheta_i}=1$, and $A_i=d$ with probability $p^{d,NE}_{\vartheta_i}=0$.

For consumers whose types are in the set $\Sigma_{2,2}$, a mixed-strategy NE with the ES policy exists if and only if the following condition holds:

 \vspace{-0.1in} 
 \small
\begin{equation}\label{eq:relation_E_0_E_1_es_ne_extra_demand}
(\gamma-\varepsilon_{\vartheta}\beta)\cdot E_{\vartheta} = (\gamma-\varepsilon_{\tilde{\vartheta} }\beta)\cdot E_{\tilde{\vartheta}} , \ \forall \vartheta , \tilde{\vartheta} \in \Sigma_{2,2}.
\end{equation}
\normalsize

To derive condition \eqref{eq:relation_E_0_E_1_es_ne_extra_demand} we re-write \eqref{eq:condition_ES_NE} first with assuming that a consumer $i$ of type $\vartheta_i \in \Sigma_{2,2}$ plays the strategy $A_i=d$ with probability $p^{d,NE}_{\vartheta_i}=1$ (in \eqref{eq:probrelation1es}) and second with assuming that a consumer $j$ with type $\vartheta_j \in \Sigma_{2,2} \setminus \{\vartheta_i\}$ plays the strategy $A_j=d$ with probability $p^{d,NE}_{\vartheta_j}=1$ (in \eqref{eq:probrelation2es}).

\vspace{-0.1in}
\begin{small}
\begin{align} \label{eq:probrelation1es}
  D^{Total}_{\Sigma_1}+1+\sum_{ {\vartheta'}\in \Sigma_{2,2}} r_{\vartheta'}~ (N-1)~p^{d,NE}_{\vartheta'}=\frac{\mathcal{RE}(\gamma-1)}{E_{\vartheta_i}(\gamma-\varepsilon_{\vartheta_i}\beta)},
\end{align}
\end{small}
\vspace{-0.1in}

\begin{small}
\begin{align} \label{eq:probrelation2es}
  D^{Total}_{\Sigma_1}+1+\sum_{ {\vartheta'}\in \Sigma_{2,2}} r_{\vartheta'}~ (N-1)~p^{d,NE}_{\vartheta'}=\frac{\mathcal{RE}(\gamma-1)}{E_{\vartheta_j}(\gamma-\varepsilon_{\vartheta_j}\beta)}.
\end{align}
\end{small}
Then, since the right-hand sides of \eqref{eq:probrelation1es}-\eqref{eq:probrelation2es} are equal, the left-hand sides will be also equal and \eqref{eq:relation_E_0_E_1_es_ne_extra_demand} derives.

Additionally, for the consumers of type $\vartheta \in \Sigma_{2,2}$, the competing probabilities that lead to NE states lie in the range $p^{min}_{\vartheta} \leq p^{d,NE}_{{\vartheta}} \leq p^{max}_{\vartheta}$, where:

 \vspace{-0.1in} 
 \footnotesize
\begin{align}
&  p^{min}_{\vartheta}=\nonumber\\&\max \left\{0,\frac{\frac{\mathcal{RE}(\gamma-1)}{E_{\vartheta}(\gamma-\varepsilon_{\vartheta}\beta)}-
  \sum\limits_{\tilde{\vartheta} \in \Sigma_{2,2} \cup \Sigma_1 \setminus \{\vartheta\}}N r_{\tilde{\vartheta}} }{N r_{\vartheta}} \right\}, \label{eq:plminbound_appendix}\\
& p^{max}_{\vartheta} = \min \left\{1,\frac{\frac{\mathcal{RE}(\gamma-1)}{E_{\vartheta}(\gamma-\varepsilon_{\vartheta}\beta)}-\sum\limits_{\tilde{\vartheta} \in  \Sigma_1 }N r_{\tilde{\vartheta}} }{N r_{\vartheta}}\right\}.
    \label{eq:plmaxbound_appendix}
\end{align}
\normalsize  
\vspace{-0.1in}

To derive the probability bounds, we re-write \eqref{eq:condition_ES_NE} assuming that all consumers of the same type play the same mixed strategy, i.e., 

\vspace{-0.1in}
\begin{small}
\begin{align} \label{eq:probrelation3es}
  D^{Total}_{\Sigma_1}+\sum_{ {\vartheta'}\in \Sigma_{2,2}} N~r_{\vartheta'}~p^{d,NE}_{\vartheta'}=\frac{\mathcal{RE}(\gamma-1)}{E_{\vartheta_i}(\gamma-\varepsilon_{\vartheta_i}\beta)}.
\end{align}
\end{small}

The minimum bound on the probability for playing RES, $p_{\vartheta}^{\min}$, derives by setting in (\ref{eq:probrelation3es}) $p^{d,NE}_{\tilde{\vartheta}}=1$, $\forall \tilde{\vartheta}\in \Sigma_{2,2}$ with $\tilde{\vartheta}\neq \vartheta=\vartheta_i$. Similarly, the maximum bound on the probability for playing RES, $p_{\vartheta}^{\max}$, derives by setting in (\ref{eq:probrelation3es}) $p^{d,NE}_{\tilde{\vartheta}}=0$, $\forall \tilde{\vartheta}\in \Sigma_{2,2}$ with $\tilde{\vartheta}\neq \vartheta=\vartheta_i$. 

The Remarks 3 and 4, which are stated for the PA allocation policy in Section \ref{sec:gameanalysis}, also hold in case of the ES allocation policy. 

The social cost under the ES policy can be expressed as 

\footnotesize
\begin{align}
&C^{ES}(\mathbf{p^{NE}}) =  N \sum_{\vartheta \in \Theta} r_{\vartheta}~ \min\{sh(\mathbf{p^{NE}}), E_{\vartheta}\} ~p_{\vartheta}^{d,NE}~ c^{RES} \nonumber\\ + &\left[D(\mathbf{p^{NE}}) -N \sum_{\vartheta \in \Theta} r_{\vartheta} ~\min\{sh(\mathbf{p^{NE}}), E_{\vartheta}\} ~p_{ \vartheta}^{d,NE}\right]
 ~c^{grid,d}\nonumber\\ + &
 N \left[ \sum_{\vartheta \in \Theta} r_{\vartheta }~ p^{n,NE}_{\vartheta}~ \varepsilon_{\vartheta }~E_{\vartheta}\right] c^{grid,n}.
 \label{eq:social_cost_es_sc}
 \end{align}
\normalsize

\subsection{Centralized Energy Sharing Mechanism Under ES Policy}
Similar to C-ESM under the PA policy (Section \ref{sec:coordinated}), the C-ESM under the ES policy is modeled as an optimization problem, defined as:

 \vspace{-0.1in} 
 \begin{small}
 \begin{subequations} \label{eq:social_cost_x_opt_es}
\begin{alignat}{2}
& \min_{\mathbf{p}} \ && C^{ES}(\mathbf{p}) \label{eq:opt_1_es} \\
 & \text{s.t. } &&p^{d}_{\vartheta},~ p^{n}_{\vartheta}\geq 0  ,  \ \forall \vartheta \in \Theta \label{eq:opt_2_es} \\
 & \quad && p^{d}_{\vartheta} + p^{n}_{\vartheta} = 1,  \ \forall \vartheta \in \Theta. \label{eq:opt_4_es}
 \end{alignat}
 \end{subequations}
\end{small} 

The problem \eqref{eq:social_cost_x_opt_es} is non-convex due to its objective function and the form of the equal share $sh(\mathbf{p^{NE}})$ (Eq. \eqref{eq:fairshare}). In our simulations in Section \ref{sec:comptoES}, we solve it with genetic algorithms using the Global Optimization Toolbox of MATLAB. 